\documentclass[prx,aps,twocolumn,superscriptaddress,showpacs,floatfix,amssymb,amsmath]{revtex4-1}
\usepackage{epsfig}
\usepackage{graphics}

\usepackage{float} 
\usepackage{array}
\usepackage{multirow}
\usepackage{combelow}
\usepackage{yfonts}
\usepackage{graphicx}
\usepackage{amssymb}
\usepackage{mathrsfs}
\usepackage{bbm}
\usepackage{epsfig}
\usepackage{makecell}

\usepackage[usenames,dvipsnames]{color}

\usepackage{bbm}
\usepackage{color}

\newcommand{\be}{\begin{eqnarray}}
\newcommand{\ee}{\end{eqnarray}}

\usepackage{xcolor}
\usepackage{ulem}
\definecolor{purple}{rgb}{0.8,0,0.6}

\allowdisplaybreaks

\begin{document}


\title{
\Large%
Local topology and perestroikas in protein \\ structure and folding dynamics
}

\author{Alexander Begun}
\email{beg.alex93@gmail.com}
\affiliation{Nordita, Stockholm University, Roslagstullsbacken 23, SE-106 91 Stockholm, Sweden}

\author{Maxim N.~Chernodub}
\email{maxim.chernodub@univ-tours.fr}
\affiliation{Institut Denis Poisson, CNRS UMR 7013, Universit\'e de Tours, 37200 France}
\affiliation{Department of Physics, West University of Timi\cb{s}oara, \\
Bd.~Vasile P\^arvan 4, Timi\cb{s}oara 300223, Romania}
\author{Alexander Molochkov}
\email{molochkov.alexander@gmail.com}
\affiliation{Pacific Quantum Center, Far Eastern Federal University, 690950, Sukhanova 8, Vladivostok, Russia}
\author{Antti J. Niemi}
\email{Antti.Niemi@su.se}
\affiliation{Nordita, Stockholm University, Roslagstullsbacken 23, SE-106 91 Stockholm, Sweden}

\begin{abstract}
{

Methods of local topology are introduced to the field of protein physics. This is achieved by explaining how the 
folding and unfolding processes of a globular protein alter the local topology of the protein's C$\alpha$ backbone 
through conformational bifurcations. The mathematical formulation builds on  the concept of 
Arnol'd's perestroikas,  by extending it to piecewise linear chains using the 
discrete Frenet frame formalism. In the low-temperature folded phase, the backbone geometry generalizes the concept 
of a Peano curve, with its modular building blocks modeled by soliton solutions of a discretized nonlinear Schr\"odinger 
equation. The onset of thermal unfolding begins when perestroikas change the flattening and branch points that determine 
the centers of solitons. When temperature increases, the perestroikas cascade, which leads to a progressive disintegration 
of the modular structures. The folding and unfolding processes are quantitatively characterized by a correlation function 
that describes the evolution of perestroikas under temperature changes. The approach provides a comprehensive 
framework for understanding the Physics of protein folding and unfolding transitions, contributing to the broader field 
of protein structure and dynamics. 
}
%
\end{abstract}
\maketitle

\vfill\eject

\section{Introduction}

Local topology aims to understand the local and small-scale properties of a space in terms of the immediate 
surroundings of its points, using the rules and concepts of topology. Bifurcation theory is designed to describe 
phenomena where a small continuous change in the parameter values of a system can cause a sudden 
change in its local topological structure. Both local topology and bifurcation theory have a significant role in the 
mathematical study of dynamical systems and differential equations, with applications that range from 
chaos theory and phase transitions to chemical reactions and evolutions between different states of 
biological function~\cite{Strogatz}.

At a technical level, the way how a bifurcation changes the local topology of a system is exemplified by the 
following elemental evolution equation:
\begin{equation}
 \frac{d\varphi(t)}{dt} \ \equiv \ \varphi_t  =   m \varphi  - \varphi^3\,. 
\label{saddle}
\end{equation}
Here $\varphi(t)$ is a real valued function that describes the state of a system as a function of the variable $t$, 
and $m$ is called the bifurcation 
parameter that acts as a control parameter. When $m>0$, there is an unstable fixed point at $\varphi=0$ 
and two stable fixed points at $\varphi = \pm \sqrt{m}$. As $m\to 0$, the system undergoes a supercritical 
pitchfork bifurcation, causing the fixed points to coalesce so that for $m<0$, there exists only a single stable 
fixed point at $\varphi=0$. In this way, by changing the number and character of its fixed points,  the bifurcation 
also changes the local topology of the system. Familiar physical scenarios where a pitchfork bifurcation 
described by an equation such as (\ref{saddle}) can be encountered include the Landau theory of 
superconducting and ferromagnetic phase transitions. 

Directly relevant to us in the sequel, the following example provides a more elaborate physical example of the effect of the 
pitchfork bifurcation. Here, the state of the system varies locally according to a function $\varphi(s)$ that solves
 the equation.
\begin{equation}
 \frac{d^2 \varphi}{ds^2} \ \equiv \ \varphi_{ss}  =  \varphi^3 - m\varphi\,, 
 \label{kink1}
 \end{equation}
describing the critical points of the free energy 
\begin{equation}
F(\varphi) = \int \! ds  \left\{ \frac{1}{2} (\varphi_s)^2 + \frac{1}{4} (\varphi^2 - m )^2 \right\}\,.
\label{Fkink}
\end{equation}

\begin{figure}[h]
        \centering
                \includegraphics[width=0.45\textwidth]{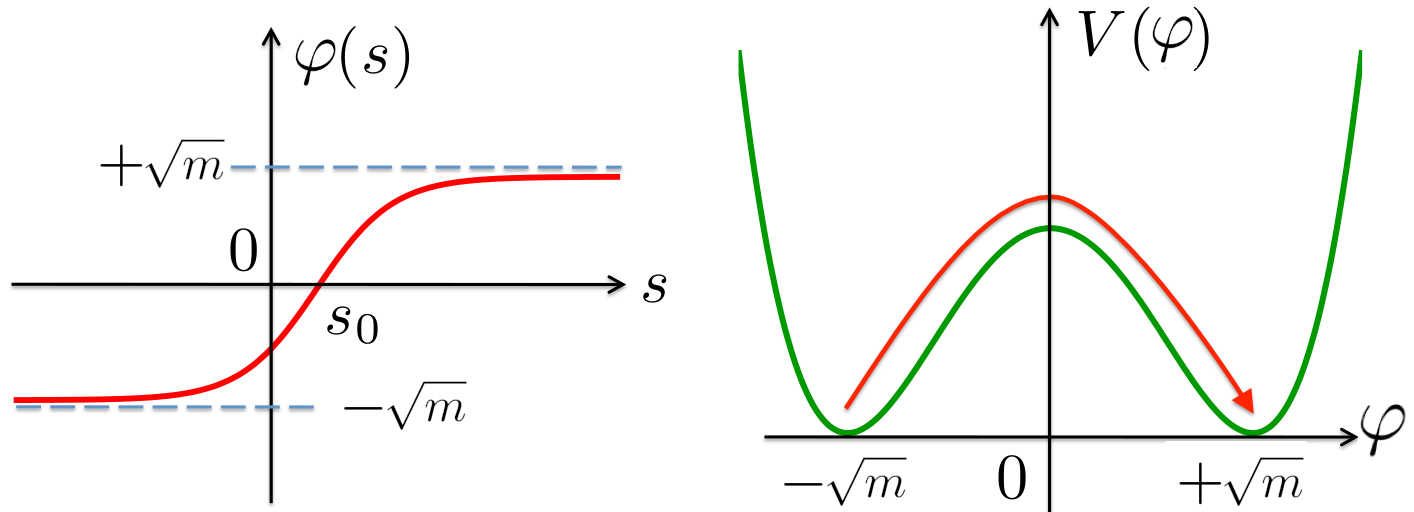}
        \caption{{
       The soliton (\ref{kink2}) is a topologically stable domain wall that interpolates between the two ground 
       states at $\varphi = \pm\sqrt{m}$ of the potential $V(\varphi)$ of (\ref{Fkink}) as $s \to \pm\infty$. 
          }}
       \label{fig1}
\end{figure}

When the control parameter $m>0$, the  equation (\ref{kink1})  is  solved by the following topological soliton~\cite{Manton}:
 \begin{equation}
 \varphi(s) = \pm \sqrt{m} \tanh \left[ \sqrt{ \frac{m}{2} } (s-s_0) \right]\,. 
\label{kink2}
\end{equation}
It has the profile of a domain wall that interpolates between the two attractive fixed points of (\ref{saddle}), {\it i.e.} 
minima of the potential in Eq.~(\ref{Fkink}), as shown in Figure~\ref{fig1}. Its topological stability derives from the 
boundary conditions at $s\to\pm\infty$ that can not be changed by any local, finite energy deformation of $\varphi(s)$.
 %
 %
 %
 %
 %
 %
 %
 %
 %

When $m\to 0$ the pitchfork bifurcation takes place, and  for $m<0$ a single topological soliton can no longer
exist.  

Proteins are examples of linear polymers that can undergo such changes in their local topology, often with 
important biological ramifications.  Here, we introduce and develop the relevant concepts of local topology 
and bifurcation theory to explain how proteins fold by transiting between different phases, from random chains
to biologically active and structured conformations. More generally, we propose to use local topology and 
bifurcation theory to understand the dynamics and the physical functioning of proteins within living 
organisms: While artificial intelligence algorithms such as AlphaFold \cite{alpha} are highly successful 
in predicting the final folded protein structures, thus far, these approaches have not been extended to 
describe the dynamical aspects and details how protein folding progresses. We trust that by revealing 
deep connections between  protein (un)folding, local topology and bifurcation dynamics, 
we also provide an impetus to develop machine-learning approaches that model protein function. 

Our starting point is Arnol'd's generalization of bifurcation theory. In a series of seminal 
articles~\cite{arnold1991number,arnold1995number,arnold1996number} Arnol'd extended bifurcation 
theory to elucidate complex and intricate changes that can take place in the local topology of function 
families as their parameters vary. Central to his approach is the concept of a `perestroika' -- a term that 
he coined to describe certain topological reorganizations that can occur in complex systems that are more 
general than those governed by canonical bifurcation theory. Specifically, he categorized those perestroikas 
that influence local topology changes in the case of smooth space 
curves~\cite{arnold1996number,aicardi2000self,uribe2004singularities}. 

We shall adapt Arnol'd's perestroikas to describe protein folding and unfolding transitions. However, since 
the C$\alpha$  backbone of a protein is a piecewise linear polygonal chain rather than a continuously 
differentiable function, we first need to extend and modify Arnol'd's framework, so that we can 
address local topology 
and its changes by perestroikas in the case of discrete chains.

The literature \cite{degen,schafer} usually assigns four distinct structural states, or phases, to a linear 
heteropolymer such as the protein C$\alpha$ backbone.  The details of a phase diagram and the conditions 
that determine the phase where a given C$\alpha$ backbone resides depend on many factors, such as chemical 
composition,  temperature,  pressure, and the quality of solvent. Each of the different phases has its 
own universal geometric characteristics that are grounded in fractal geometry and considered to be largely 
independent of the atomic level details of the protein. To determine the phase of a given C$\alpha$ backbone, 
one commonly inspects its radius of gyration $R_g$ as a reaction coordinate, also called an order parameter in 
statistical physics. In the general case of a discrete point set $\mathbf r_i$ ($i=1,\dots, N$) such as the vertices of a 
discrete piecewise linear chain, the radius of gyration is 
\begin{equation}
R_g \ = \ \sqrt{ \, \frac{1}{2N^2}  \sum_{i,j=1}^N ( {\bf r}_i  - {\bf r}_j )^2\, }\,.  
\label{Rg}
\end{equation}
Its scaling properties determine the phase of the chain  as follows: In the limit where the number of vertices $N$ is 
very large, $R_g$  has the asymptotic expansion \cite{schafer,lim}
\begin{align}
R_g  \ \buildrel{N \ {\rm large}}\over{\xrightarrow{\hspace{12mm}}}  
& \ R_0 N^{\nu} ( 1 + R_1 N^{-\delta_1} + \dots  )  \nonumber \\
& \hskip -5mm \sim \ R_0 N^{\nu} \ + \ \dots\,.
\label{R}
\end{align}
The exponent $\nu$  governs the large-$N$ asymptotic,  the Kuhn length $R_0$ is the average distance between 
the neighboring vertices, and the $R_1$, $\delta_1$ {\it etc.} characterize finite-size corrections. The numerical 
values of the exponents $\nu$ and $\delta_1$ {\it etc.} have a universal character that is independent of the atomic 
level structure~\cite{degen,schafer,nickel,lim}, and often $\nu$ is also interpreted as the inverse Hausdorff dimension 
of the backbone even though the two measure somewhat different properties of a point set.  On the other hand, 
the pre-factors $R_0$, $R_1$, {\it etc.} are non-universal quantities, with values that 
can, in principle, be computed from atomic-level details such as temperature, pressure, and chemical composition 
of the solvent. 

In the case of a given protein C$\alpha$ backbone the  number of amino acids $N$ is fixed, so that 
the limit (\ref{R}) can not be 
directly considered. However, one can still estimate the value of $\nu$ \cite{Huang} by comparing the radius of gyration 
of a given protein to statistical average values in the Protein Data Bank (PDB) \cite{PDB}.  
The following mean field values are commonly assigned to the value of $\nu$, in the case of a protein C$\alpha$ 
backbone~\cite{degen,schafer,nickel,lim,Huang}: 
\begin{equation}
 \nu \ = \  \left\{  \ \, \begin{matrix} 1/3  \\ 1/2  \\ 3/5   \\ 1 \end{matrix} \right.  \ \ \ \ \  \begin{matrix}
\  {\rm folded}  \\ 
 {\rm molten ~ globule}  \\  {\rm self{-}avoiding}  \\  {\rm quasilinear}  \end{matrix} \,
\label{nuval}
\end{equation}
Under poor solvent conditions or at low temperatures, a protein collapses into a  folded conformation
 \cite{huggins,flory} that can be characterized by the mean-field value $\nu = 1/3$. Biologically active globular proteins 
 are often assigned this value. Notably,  its inverse coincides with the Hausdorff dimension $D_H=3$ of space-filling curves such as 
 the three-dimensional Peano curve and its various generalizations~\cite{Sagan-2012}. When the ambient conditions 
 are such that the attractive and repulsive forces between different backbone segments become balanced, the protein 
 resides in the molten globule phase, which is characterized by the mean-field value $\nu =1/2$. This is also the 
 inverse value of the Hausdorff dimension  in the case of a three-dimensional Brownian random walk, even though 
 the two are  different geometrical structures as we shall demonstrate.  
 The molten globule separates the space-filling folded phase from the fully unfolded, high-temperature self-avoiding 
 random walk (SARW) phase for which the mean field Flory value $\nu = 3/5$ is found. Finally, when $\nu =1$, the 
 backbone chain loses its inherently fractal structure and adapts to a quasilinear, filamental geometry. Biologically 
 active collagen, which is the most abundant protein in mammals, provides an example of a protein that resides in 
 this phase. It has also been argued that this value of $\nu$ relates to the phenomenon of cold denaturation of a 
 protein~\cite{Privalov}.

We begin with a concise survey of Arnol'd's perestroikas in the case of differentiable curves. We then 
proceed to extend this mathematical formalism to piecewise linear chains, which is the case that is relevant 
to protein C$\alpha$ backbones. Using a crystallographic myoglobin structure as a representative 
example of a folded globular protein, we analyze the local topology and bifurcations that govern its folding and 
unfolding dynamics. For quantitative analysis, we introduce a novel correlation function that describes 
how bifurcation dynamics proceed as the ambient temperature changes. Since our focus lies on topological 
characteristics that are quite universal and insensitive to detailed backbone geometry that depends on 
specifics of chemical composition, the insights that we gain from the myoglobin case study can be extended
intactum to other globular proteins and, more broadly, to linear polymers.

\section{Methods}

\subsection{The Frenet-Serret frames}

%

%
We begin with a brief review of the standard differential geometry of curves in $\mathbb R^3$, a classical subject 
commonly found in textbooks, for example~\cite{spivak1970comprehensive}. We consider a length-$L$ space curve 
$\mathbf x(s)$ with $s\in [s_A , s_B]$ the arc-length parameter so that $s_B - s_A = L$ and $|| \dot{\mathbf x}(s) || = 1$. 
We assume that the curve does not self-intersect, and for technical reasons, we take $\mathbf x(s)$ to be at least 
four times continuously differentiable if not smooth. We focus on local aspects so that the curve can be open or closed, 
the shape is generic, and it can change freely by local deformations. We are interested in local topological invariants, 
those properties of the curve that are robust against smooth shape deformations when we observe the curve at a small enough scale.

By shifting each point $\mathbf x(s)$ of the curve 
by a very short distance $\boldsymbol \epsilon(s)$ at that point, we obtain a framing of the curve. The self-linking 
number of $\mathbf x(s)$ is a quantity that measures how many times the curve $\mathbf x(s)$ winds, or links, around its 
shifted version $\mathbf x(s) + \boldsymbol \epsilon(s)$. Since the self-linking number takes into account the entire structure 
of the curve rather than just the properties of a small neighborhood around any point on the curve, it appears to be more like 
a global than a local topological invariant. But the value of the self-linking number depends on the choice of the framing 
vector $\boldsymbol \epsilon(s)$, and if we allow $\boldsymbol \epsilon(s)$ to vanish at a point along the curve, the 
self-linking number can change. The framing then becomes a local topological invariant, that remains intact only 
under those continuous changes that retain $\boldsymbol \epsilon(s)$ non-vanishing.

%
In classical differential geometry of a curve, the framing is commonly determined by Frenet frames~\cite{spivak1970comprehensive}. 
At a generic point $\mathbf x(s)$ of the curve, the right-handed orthonormal Frenet frame is defined by three 
vectors, the unit length tangent vector:
\begin{equation}
\mathbf t =  \frac{d \mathbf x(s)}{ds} \equiv \dot {\mathbf x} (s)\,,
\label{tc}
\end{equation}
the unit length binormal vector
\begin{equation}
\mathbf b =  \frac{ \dot{\mathbf x}\times \ddot{\mathbf x} }{|| \dot{\mathbf x} \times \ddot{\mathbf x} || } \,,
\label{bc}
\end{equation}
and the unit length normal vector
\begin{equation}
\mathbf n = \mathbf b \times \mathbf t\,,
\label{nc}
\end{equation}
and we may choose either $ \mathbf  b(s)$ or $\mathbf n(s)$ or their linear combination 
$\boldsymbol \epsilon(s)  = \varepsilon_1 \mathbf b + \varepsilon_2 \mathbf n$ as a framing vector to promote  
the curve $\mathbf x(s)$ into a framed curve. The ambient geometry of the curve is then determined uniquely 
by two scalar functions. The first function is the positive valued curvature, 
\begin{equation}
\varkappa(s) \ = \ \frac{ || \dot {\mathbf x} \times \ddot {\mathbf x} || } { || \dot  {\mathbf x} ||^3 }\,.
\label{kappa}
\end{equation}
It is inversely proportional to the radius of the osculating circle, describing how the curve bends along its 
osculating plane.  The second function is the real-valued torsion $\vartheta(s)$
\begin{equation}
\vartheta(s) \ = \ \frac{ (\dot {\mathbf x} \times \ddot {\mathbf x}) \cdot {\dddot {\mathbf x} }} { || \dot 
{\mathbf x} \times \ddot {\mathbf x} ||^2 }\,.
\label{vartheta}
\end{equation}
It measures the rate at which the curve twists {\it i.e.} rotates out of its osculating plane. For a curve 
with a non-vanishing curvature $\varkappa(s) >0$, the interrelations between the framing and the 
geometry are summarized  by the Frenet-Serret equation~\cite{spivak1970comprehensive}:
\begin{equation}
\frac{d}{ds} \left( \begin{matrix} \mathbf n \\ \mathbf b \\ \mathbf t \end{matrix} \right) \ = \ 
 \left( \begin{matrix} 0 & \vartheta & -\varkappa \\ -\vartheta & 0 & 0 \\
 \varkappa & 0 & 0  \end{matrix} \right) \left( \begin{matrix} \mathbf n \\ \mathbf b \\ \mathbf t \end{matrix} \right) \,. 
\label{frene}
\end{equation}
The fundamental theorem of space curves states that the solution of (\ref{frene}) defines the curve 
completely and uniquely, up to a global rotation and a global translation. 

%
%
%
%
%
%
%
%
%

\subsection{Inflection points}
\label{sectIP}

At an inflection point where $\varkappa(s)=0$, the Frenet frames can not be determined. For that reason, 
differential geometry textbooks~\cite{spivak1970comprehensive}  commonly assume that the curvature 
(\ref{kappa}) is non-vanishing, $\varkappa(s)>0$. Since we are interested in the local topology of the curve and, 
in particular, those continuous shape variations that can change it, we need to develop an appropriate extension of the
 traditional textbook presentation of the Frenet frame formalism to include inflection points where $\varkappa(s)$ vanishes. 

Thus, we consider a curve with $N$ isolated and non-degenerate inflection points $\mathbf x(s_n)$ with 
parameter values $s=s_n$ ($n=1,\dots ,N$) between the endpoints $s_A$ and $s_B$:
\[
s_A < s_1 <  \dots  < s_N < s_B\,.
\] 
For clarity, we assume that in the limit, when we approach an inflection point $s_n$ from either direction 
along the curve $s\to s_n^{\pm}$, the left and the right derivatives of $\varkappa(s)$ are non-vanishing 
and equal in magnitude but with opposite signs:
\[
\lim\limits_{s\to s_n ^- } \frac{ d \varkappa (s)}{ds} = - \lim\limits_{s\to s_n^+ } \frac{d \varkappa (s) }{ds} \ \not= 0 \,.
\]
We may then extend the piecewise-defined non-negative geometric curvature $\varkappa(s)$ into a 
real-valued, signed curvature $\kappa(s)$ that is defined on the entire segment $s\in [s_A,s_B]$ as a 
continuously differentiable function, 
\begin{equation}
 \kappa(s) \ = \  \varkappa(s) \exp\Bigl\{ i \pi \sum_{n=1}^N (-)^{n+1} \theta(s-s_n) \Bigr\}\,,
\label{tilde}
\end{equation}
where $\theta(s)$ the Heaviside step-function. In particular, at non-degenerate inflection points, one finds:
\[
\frac{ d \kappa }{ds}{\biggl |}_{{s = s_n}}  \not= 0 \,.
\]
To deduce the effect of the presence of the inflection point on torsion, we start with a general rotation of 
$\mathbf n(s)$ and $\mathbf b(s)$ around $\mathbf t(s)$ by an arbitrary angle $\eta(s)$. The result is another 
orthonormal pair $\mathbf e_1(s), \mathbf e_2(s)$:
\begin{equation}
\left( \begin{matrix} {\bf n} \\ {\bf b} \end{matrix} \right) \ \to \
\left( \begin{matrix} \cos \eta(s) & - \sin \eta(s) \\ \sin \eta(s) & \cos \eta(s) \end{matrix}\right)
\left( \begin{matrix} {\bf n} \\ {\bf b} \end{matrix} \right) \ \equiv \  \left( \begin{matrix} {{\bf e}_1} \\ {\bf e}_2 
\end{matrix} \right) \,. \
\label{newframe}
\end{equation}
This transformation converts the original Frenet-Serret equation into 
\begin{equation}
\frac{d}{ds} \left( \begin{matrix} {\bf e}_1 \\ {\bf e }_2 \\ {\bf t} \end{matrix}
\right) =
\left( \begin{matrix} 0 & (\vartheta - \dot \eta) & - \varkappa \cos \eta \\ 
- (\vartheta - \dot \eta)  & 0 & -\varkappa \sin \eta \\
\varkappa \cos \eta & \varkappa \sin \eta  & 0 \end{matrix} \right)  
\left( \begin{matrix} {\bf e}_1  \\ {\bf e }_2 \\ {\bf t} \end{matrix}
\right) \,.
\label{contso2}
\end{equation} 
Comparison of (\ref{tilde}) and (\ref{contso2}) shows that a transition from the non-negative, geometric 
curvature $\varkappa(s)$ to the continuously differentiable signed curvature $\kappa(s)$ sends the 
torsion $\vartheta(s)$ into 
\begin{align}
\tau (s) & \ = \ \vartheta(s)  -  \dot \eta(s) \ = \ \vartheta(s) +  \pi  \frac{d} {ds} \sum_{n=1}^{N} (-)^n \theta(s-s_n) \nonumber \\
& \ = \ \vartheta(s) + \pi  \sum_{n=1}^{N} (-)^n \delta(s-s_n)\,.
\label{contgau}
\end{align}
Thus, whenever the curve passes an inflection point, the Frenet framing undergoes an instantaneous rotation 
by an angle $\pm \pi$ around its tangent vector $\mathbf t(s)$, with the choice of sign determined by the sign of torsion.  
In particular, the self-linking number determined by Frenet framing is a local topological invariant. Its value changes 
under those shape changes that engage inflection points. 

Finally,  in the generic frame Frenet equation (\ref{contso2}), we identify the gauge structure of an Abelian 
Higgs model, widely encountered in high energy and condensed matter physics.
This structure comes out when we combine the positive valued Frenet curvature $\varkappa(s)$ 
together with the angle $\eta(s)$ into a complex-valued function akin to the Higgs scalar and extend the 
torsion $\vartheta(s)$ similarly into an analogy of
an Abelian gauge connection as follows, 
\begin{equation}  
\varkappa(s) \ \to \ e^{i\eta(s)} \varkappa(s) \ \ \ \ \ \ \& \ \ \ \ \ \ \vartheta(s) \ \to \ \vartheta(s) - \frac{d \eta(s)}{ds}\,.
\label{U(1)}
\end{equation}
In this parlance, a frame rotation corresponds to a U(1) gauge transformation, with the Frenet framing 
corresponding to the unitary gauge in the Abelian Higgs model. In particular, the following combination of 
Frenet frame curvature and torsion remains intact under frame rotations,
\begin{equation} 
\phi(s) = \varkappa (s) \exp\Bigl\{  i \int_{s_A}^s ds^\prime \, \vartheta(s^\prime) \Bigr\}\,.
\label{phivar}
\end{equation}

\subsection{Local topology and perestroikas}
\label{sectFSp}

There are circumstances when even small changes in the shape of a curve can alter its local topology. 
This effect occurs when a shape change gives rise to a bifurcation that Arnol'd called a perestroika
\cite{arnold1991number,arnold1995number,arnold1996number,aicardi2000self,uribe2004singularities}. 
In particular, he showed that in the case of one-parameter families of shape-changing curves, there are 
only two perestroikas where the local topology of a generic curve can change. These were called the inflection 
point perestroika and the bi-flattening perestroika by Arnol'd. To identify and describe them, we use the Frenet 
equation to Taylor-expand the curve coordinate around an ordinary point $\mathbf x(s)$, where we, for 
clarity, choose $s=0$. With $\mathbf t,  \varkappa, \vartheta $ {\it etc.} now denoting the corresponding 
quantities when evaluated at $s=0$, the expansion in powers of $s$ proceeds as follows:
\begin{align}
\mathbf x(s) & =   \mathbf x(0) \ +\  \mathbf t  s  \ +  \ \frac{1}{2} \varkappa \mathbf n\, s^2 \nonumber \\
&  + 
\frac{1}{6} \left( \varkappa \vartheta  \mathbf b + \varkappa_s \mathbf n - \varkappa^2 \mathbf t \right)  s^3 \nonumber\\
& +  \frac{1}{24}  \left[  ( 2    \varkappa_s \vartheta + \varkappa \vartheta_s  ) \mathbf b 
   +  (\varkappa_{ss} - \varkappa \vartheta^2 - \varkappa^3 )  \mathbf n \right. \nonumber\\
   & \left. -  \ 3 \varkappa \varkappa_s \mathbf t \right]  s^4 \nonumber\\
   & +   \frac{1}{120}  \bigl[   \left(  \varkappa  \vartheta_{ss}   + 3 \varkappa_s   \vartheta_s  + 3  \varkappa_{ss} \vartheta  
    - \varkappa^3 (  \varkappa \vartheta^3 + \vartheta)  \right) \mathbf b \nonumber\\
   & + \left( \varkappa^3 - 3 \varkappa_s \vartheta^2
    - 6 \varkappa^2 \varkappa_s  - 3 \varkappa \vartheta \vartheta_s  \right) \mathbf n \nonumber\\
   & +  \left(  \varkappa^2 \vartheta^2 + \varkappa^4 - 3 \varkappa_s^2 - 4 \varkappa  \varkappa_{ss} 
   \right) \mathbf t \bigr] s^5 + \mathcal O\left(s^6\right)\,.
\label{repre}
\end{align}
The various special points of the curve where either the curvature, the torsion, or their derivatives 
vanish can then be assigned a symbol consisting of three positive integers $a_1<a_2<a_3$, with 
$a_i \in {\mathbb N}$ for  $i = 1,2,3$, and chosen to be the smallest possible natural numbers in 
powers of $s$ that represent the point in terms of the basis ($\mathbf t, \mathbf n, \mathbf b$). 
For example, in the case of an ordinary point, the symbol is $(a_1,a_2,a_3) = (1,2,3)$ since these 
are the leading powers of $s$ in (\ref{repre}) in terms of the basis vectors.

There are three special points that are of interest to us. The first is a simple inflection point, {\it i.e.} a point 
where $\varkappa=0$ but $\varkappa_s \not= 0$, and from (\ref{repre}) we read that it has the 
symbol $(1,3,4)$. Similarly, (\ref{repre}) reveals that the second is a simple flattening point with 
$\vartheta=0$ but both $\vartheta_s\not=0$ and $\varkappa \not=0$ so that the symbol is $(1,2,4)$. 
Finally, the third is a bi-flattening point with $\vartheta=\vartheta_s =0$ but $\varkappa \not=0$ and the 
symbol is $(1,2,5)$. We analyze how they affect the local topology using the expansion (\ref{repre}).

We first assume that $ \mathbf x(0)$ is a single inflection point in a parameter segment $[s_a,s_b]$ 
under consideration. From its symbol $(1,3,4)$, we conclude that the co-dimension of an inflection 
point is two. Thus, if we deform the curve around ${\mathbf x}(0)$ in a manner that retains the 
osculating plane, the inflection point can move along the curve. But if, instead, we lift the curve off its 
osculating plane the inflection point becomes removed. In particular, a generic space curve does 
not have any inflection points, and a generic one-parameter family of curves can only have isolated 
parameter values at which an inflection point appears. When this situation occurs, the curve 
undergoes a bifurcation that is called an inflection point perestroika 
\cite{arnold1991number,arnold1995number,arnold1996number} where the local topology changes. 
Specifically, the Frenet self-linking number of the curve changes in the process. 

To further illustrate the properties of inflection-point perestroika, we consider two {\it a priori} generic 
curves $\mathbf x_1(s)$ and $\mathbf x_2(s)$ with no inflection points so that they both have their 
respective Frenet framings with ensuing self-linking numbers. We also assume that the curves are in 
the vicinity of, and isotopic to, a third curve $\mathbf x(s)$ that has an inflection point $\mathbf x_0$  
with corresponding discontinuity in its Frenet framing. Furthermore, we assume that the two curves 
are located on the opposite sides of the hypersurface, which is defined by all those curves that have 
an inflection point, including $\mathbf x_0$.  Even though the two curves  $\mathbf x_1(s)$ and 
$\mathbf x_2(s)$ are mutually isotopic and located in the vicinity of each other since they are separated 
by a curve with an inflection point, their Frenet self-linking numbers are different, and any on-parameter 
family of curves that interpolates between $\mathbf x_1(s)$ and $\mathbf x_2(s)$ undergoes an 
inflection point perestroika.

Unlike an inflection point, we conclude from its symbol (1,2,4) that the co-dimension of a flattening point is 
one. Thus, a flattening point is generic, and at least one flattening point is ordinarily present along a typical 
curve. Moreover, since the torsion changes sign at a flattening point, a flattening point is a local topological 
invariant that can not be removed by small local deformations. A small local deformation of a curve can only 
transport an isolated flattening point to another place along the curve. But when the shape of a curve changes 
so that a pair of flattening points come together, they combine into a single bi-flattening point, with the 
symbol (1,2,5). This bi-flattening point can then be removed by a further, generic local deformation of the curve: 
A bi-flattening point is not a local topological invariant. Similarly, a bi-flattening point can first be created by an 
appropriate local deformation of a curve, and when the curve is further deformed, the bi-flattening point can 
become resolved into two separate flattening points. 
When either of these two events occurs, the curve undergoes a bifurcation that is called a bi-flattening 
perestroika \cite{arnold1991number,arnold1995number,arnold1996number}. 

The number of flattening points and the self-linking number that is determined by the Frenet framing 
are the only two curve-specific, mutually independent local topological invariants that can be assigned 
to a differentiable curve. Furthermore, the inflection point perestroika and the bi-flattening perestroika are 
the only two bifurcations where the number of flattening points can 
change~\cite{arnold1991number,arnold1995number,arnold1996number,aicardi2000self,uribe2004singularities}.
In the presence of an inflection point, the two 
can also interfere with each other. For example, when a curve is deformed so that two simple flattening 
points come together and disappear in a bi-flattening perestroika, the self-linking number, in general, does 
not change. But if the bi-flattening perestroika occurs in combination with an inflection point perestroika, the 
self-linking number, in general, does change.

\subsection{A limiting case}
\label{subsect:sectlimit}

We conclude our survey of differentiable curves by drawing attention to the following limit, which becomes 
relevant in the sequel of a very small but non-vanishing curvature so that
\[
\left | \frac{\varkappa(s)}{\vartheta(s)} \right | \to 0\,.
\] 
In this limit, the Frenet-Serret equation gives
\begin{equation}
\begin{aligned} 
\frac{d}{ds} (\mathbf n + i\mathbf b) & \ \approx \ &  \!\!\!\!\!\!\! \!\! - \ i \vartheta (\mathbf n + i\mathbf b)\,,  \\  
\frac{d}{ds} \mathbf t  & \  \sim \ 0\,. &   
\end{aligned}
\label{kt0}
\end{equation}
This equation describes a (almost) straight line with a framing that spirals around it at a rate and with a 
handedness that is determined by the torsion $\vartheta(s)$. At a simple flattening point the torsion 
changes its sign, and the handedness of the spiraling also changes.

%
 
 \subsection{The  discrete Frenet frames}
 
Our focus is on piecewise linear chains. 
In that case, the Frenet framing does not exist, and both the differential geometry described by the 
Frenet-Serret equation (\ref{frene}) and the concepts of local topology and perestroikas that we have 
presented need to be adapted.  The discrete Frenet frame formalism introduced in  \cite{hu2011discrete} 
provides the appropriate framework. It describes the geometry of a piecewise linear chain with vertices
$\mathbf r_i$ ($i=1,\dots ,N$). The links connecting  two neighboring vertices determine the unit tangent vectors 
\begin{equation}
\mathbf t_i = \frac{ {\bf r}_{i+1} - {\bf r}_i  }{ |  {\bf r}_{i+1} - {\bf r}_i | }\,.
\label{td}
\end{equation}
The unit binormal vectors are  
\begin{equation}
\mathbf b_i = \frac{ {\mathbf t}_{i-1} \times {\mathbf t}_i  }{  |  {\mathbf t}_{i-1} \times {\mathbf t}_i  | }\,,
\label{bd}
\end{equation}
and the unit normal vectors are 
\begin{equation}
\mathbf n_i = \mathbf b_i \times \mathbf t_i\,.
\label{nd}
\end{equation}
Together, the orthonormal triplet ($\mathbf n_i, \mathbf b_i, \mathbf t_i$) defines the discrete Frenet frame at each 
vertex $\mathbf r_i$ of the chain. 

The protein C$\alpha$ backbone structure is an important biophysical example of such a piecewise linear discrete chain. 
The vertices correspond to the positions of the C$\alpha$-atoms, and the links 
coincide with the diagonals of the peptide planes; these diagonals have a length that is very close to 3.7 \AA. 
Following  \cite{hu2011discrete}, we now outline  and  systematically tailor the formalism  of discrete 
Frenet frames for applications related to the local topology of protein C$\alpha$ backbones.

In  the case of a discrete chain, in lieu of  the continuum curvature $\varkappa(s)$ and torsion $\vartheta(s)$
the ambient geometry 
is governed by their discrete variants, the bond angles $\kappa_i$ and the torsion angles $\tau_i$. 
The values of these angles are computed from the discrete Frenet frames as follows. The bond angles are
\begin{equation}
\kappa_i \ \equiv \ \kappa_{i+1 , i} (\mathbf r_i, \mathbf r_{i+1},
\mathbf r_{i+2})  \ = \ \arccos \left( {\bf t}_{i+1} \cdot {\bf t}_i \right)
\label{bond}
\end{equation}
and the torsion angles are
\begin{align}
\tau_i & \ \equiv \ \tau_{i+1,i} ( \mathbf r_{i-1}, \mathbf r_i, \mathbf r_{i+1},
\mathbf r_{i+2}) \nonumber \\
& \ = \ {\rm sign}\{ \mathbf b_{i} \times \mathbf b_{i+1} \cdot \mathbf t_i \}
\cdot \arccos\left(  {\bf b}_{i+1} \cdot {\bf b}_i \right) \
\label{tors}
\end{align}
Notably,  the bond angle $\kappa_i$ is evaluated from three,  and  the  torsion angle $\tau_i$ is evaluated from four 
consecutive vertices. 

\begin{figure}[h]
        \centering
            \includegraphics[width=0.45\textwidth]{
            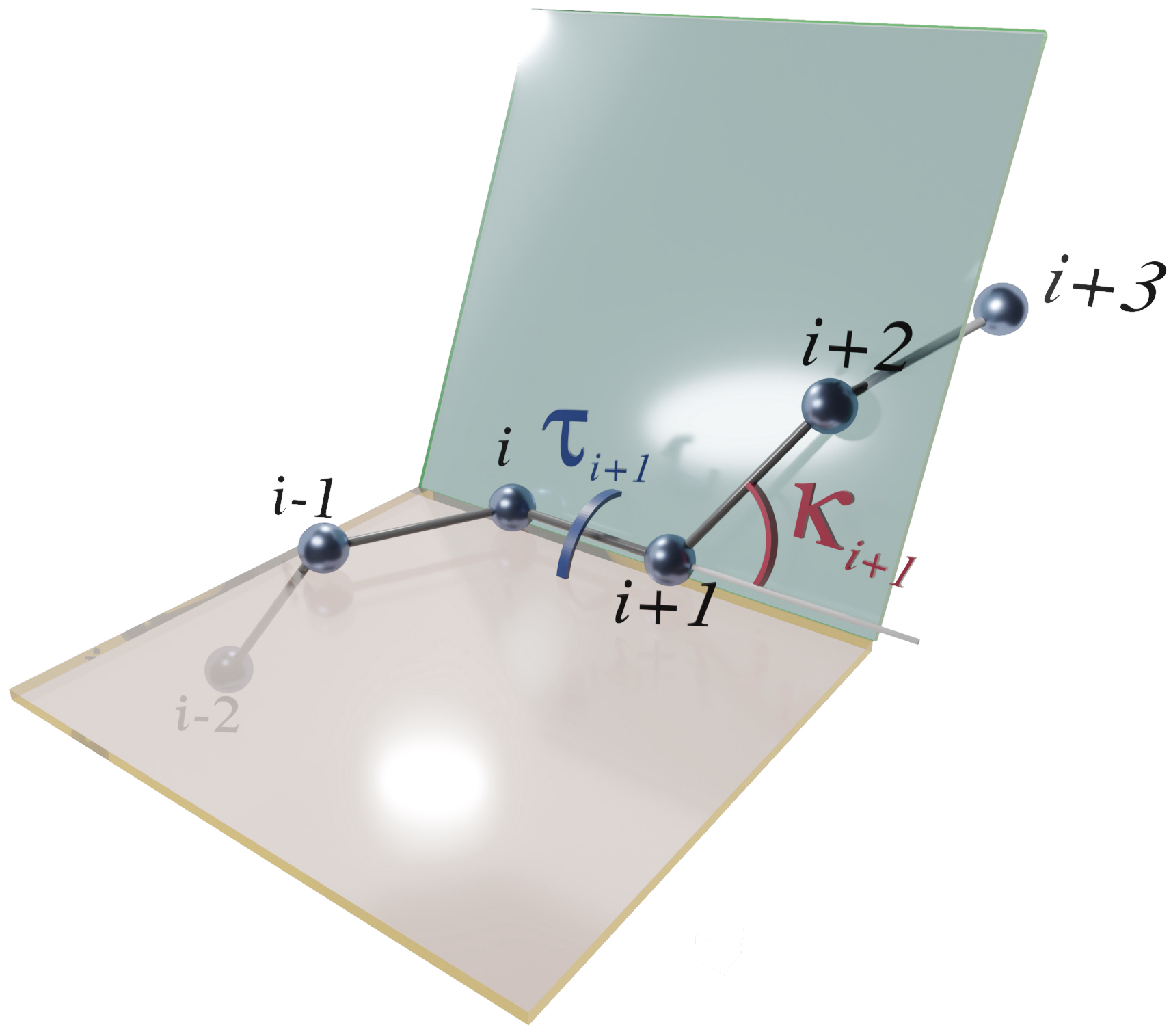}
        \caption{{
      The bond angle $\kappa_i$ is the angle between the two vectors $\mathbf r_{i+1}-\mathbf r_i$ and
    $\mathbf r_{i+2}-\mathbf r_{i+1}$. The torsion angle $\tau_i$ is the angle between the planes
    defined by vertices $\mathbf r_{i-1}, \mathbf r_i, \mathbf r_{i+1}$ and vertices  $\mathbf r_{i}, \mathbf r_{i+1}, 
    \mathbf r_{i+2}$.
            }}
       \label{fig2}
\end{figure}

Conversely, when the values of the bond and torsion angles are all known, the discrete Frenet equation~\cite{hu2011discrete}
\begin{equation}
\left( \begin{matrix} {\bf n}_{i+1} \\  {\bf b }_{i+1} \\ {\bf t}_{i+1} \end{matrix} \right)
= 
\left( \begin{matrix} \cos\kappa \cos \tau & \cos\kappa \sin\tau & -\sin\kappa \\
-\sin\tau & \cos\tau & 0 \\
\sin\kappa \cos\tau & \sin\kappa \sin\tau & \cos\kappa \end{matrix}\right)_{\hskip -0.1cm i+1 , i}
\left( \begin{matrix} {\bf n}_{i} \\  {\bf b }_{i} \\ {\bf t}_{i} \end{matrix} \right) 
\label{DFE2}
\end{equation}
computes  the discrete Frenet frame at the vertex $\mathbf r_{i+1}$ from the frame at  the preceding vertex  $\mathbf r_{i}$. 
The chain from the initial position $\mathbf r_0$ to the given vertex $\mathbf r_n$ can then be constructed 
iteratively using \cite{hu2011discrete}
\begin{equation}
\mathbf r_n \ = \ \sum\limits_{i=0}^{n-1} |\mathbf r_{i+1} - \mathbf r_i | \, \mathbf t_i
\label{chain}
\end{equation}
and in the case of a protein C$\alpha$ backbone we may set $|\mathbf r_{i+1} - \mathbf 
r_i | = 3.7$\AA. The initial vertex can be chosen as the origin $\mathbf r_0 = \mathbf 0$, the first vertex $\mathbf r_1$ can 
be placed on the positive $z$-axis and
the second vertex can be placed on the positive quadrant of the $xy$ plane. 
The Figure \ref{fig2} summarizes the discrete Frenet framing.
 %
 %
 %
 %
 %
 %
 %
 %
 %
%
%
%
%
%
%
%
Notably, in a limit where the bond length  goes to zero 
the discrete Frenet equation (\ref{DFE2}) becomes the continuum
Frenet equation  (\ref{frene}) \cite{hu2011discrete}. 

The fundamental range of a bond angle is $\kappa_i \in [0,\pi]$, and in the case of a torsion angle, 
it is $\tau_i \in [-\pi, \pi)$. Thus, these angles can be interpreted  geometrically as latitude and longitude 
angles, respectively, on a  (Frenet) two-sphere $\mathbb S^2_i$ that is centered at the $i^{th}$ vertex
$\mathbf r_i$. But,  as in the continuum case, there is  an advantage to extending the range of the bond 
angles to $\kappa_i \in [-\pi,\pi]$, while the
range of the torsion angle remains  $\tau_i \in [-\pi, \pi)$ (mod($2\pi$). This becomes substantiated when 
we observe that the following  discrete $\mathbb Z_2$ transformation  
\begin{equation}
\begin{matrix}
\ \ \ \ \ \ \ \ \ \kappa_{k+1,k} & \to  &  - \ \kappa_{k+1,k} \ \ \ \hskip 1.0cm  {\rm for \ \ all} \ \  k \geq i  \\
\ \ \ \ \ \ \ \ \ \tau_{i+1,i \ }  & \to & \tau_{i+1,i} - \pi  \hskip 1.5cm  {\rm mod}\, (2\pi)
\end{matrix}
\label{dsgau}
\end{equation}
leaves the  chain (\ref{chain}) intact.

%
 
 \subsection{Projective Frenet plane for discrete chains}

\begin{figure}[h]
        \centering
                \includegraphics[width=0.45\textwidth]{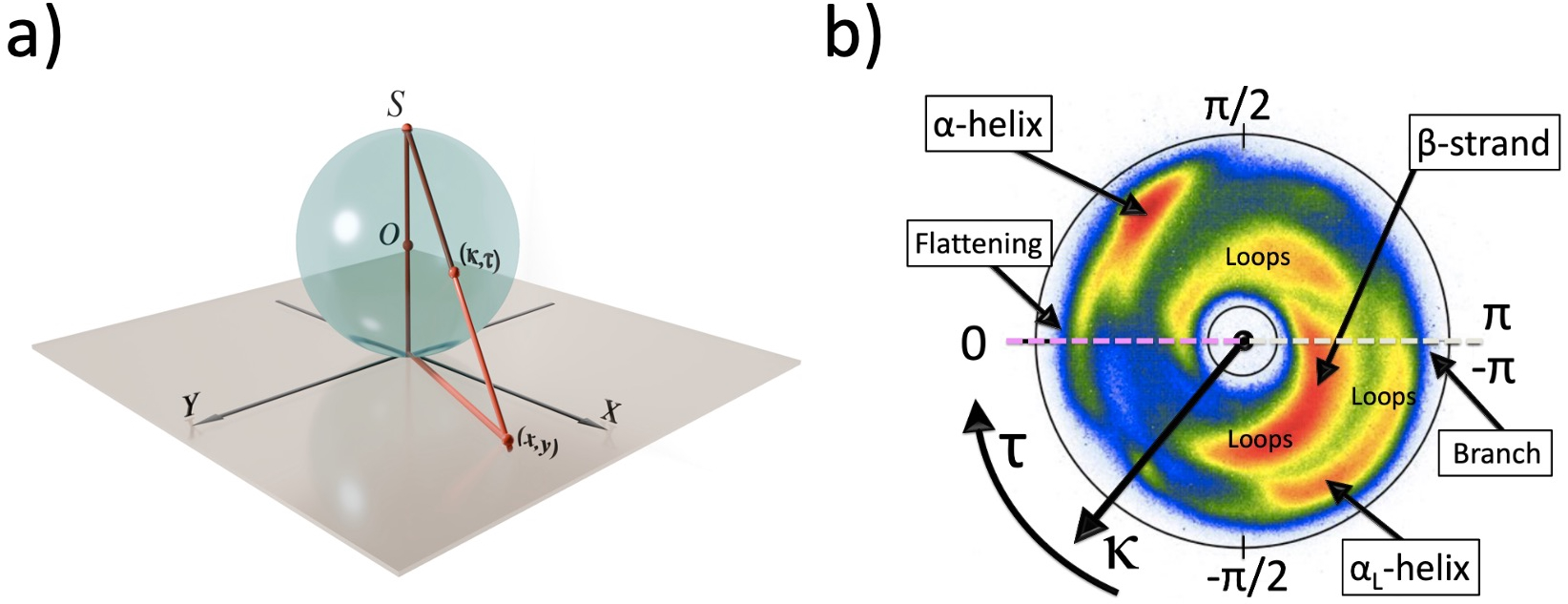}
        \caption{{a)  Stereographic projection (\ref{stere}) of  Frenet sphere to its tangent plane at the north pole from the 
        south pole $S$. b) The stereographically projected map of backbone C$\alpha$-atoms in Protein Data Bank (PDB). 
        The PDB data concentrates on an annulus $\mathbb A$ with color coding corresponding to the number of entries in 
        PDB from red large to blue small and white is none.   Major secondary structures are identified, with loops distributed
     widely over $\mathbb A$.  The stereographically projected latitude $\kappa$ measures the distance from center 
     ($\kappa=0$) and the longitude $\tau$ is angle around the center of $\mathbb A$. The line of flattening is $\tau=0$, 
     and $\tau = \pm\pi$ is the branch cut.}}
       \label{fig3}
\end{figure}

For visualization purposes, following \cite{Lundgren-2012}, we stereographically project the Frenet 
sphere $\mathbb S^2_i$ to its tangent plane at the north pole, as depicted in Figure \ref{fig3}a. For 
this, we first orient the Frenet sphere that is centered at vertex $\mathbf r_i$ so that the position 
$\mathbf r_{i+1}$ of the next vertex coincides with the north pole of $\mathbb S^2_i$ where the 
ensuing bond angle $\kappa_i$ has the value $\kappa_i= 0$.  The torsion angle  $\tau_i$ now 
measures the longitude of the sphere $\mathbb S^2_i$ so that  $\tau_i=0$ on the great circle that 
passes through the north pole and through the tip of the binormal vector $\mathbf b_i$. The 
stereographic projection of the Frenet sphere with coordinates  $(\kappa, \tau)$ to the tangent 
(Frenet) plane at the north pole with Cartesian coordinates ($x,y$)  is then computed as follows, see Figure \ref{fig3}a.
 \begin{equation}
 x + iy =  \tan  \bigl(\frac{\kappa}{2}\bigr) \cdot e^{- i \tau}\,.
 \label{stere}
 \end{equation}
%
%
%
%
%
%
%
%
%
%
%
%
%
%
%
%
%
%
%
%
%
%

We specify the vector $\mathbf t_{i+1}$  in terms of coordinates on the preceding Frenet sphere $\mathbb S^2_i$ as follows: 
We first parallel translate the vector $\mathbf t_{i+1}$ so that its base coincides with the vertex $\mathbf r_i$. We 
then record the coordinates ($\kappa_i, \tau_i$) that correspond to the tip of $\mathbf t_{i+1}$  on the surface of $\mathbb 
S^2_i$, and we stereographically project them onto the tangent plane.  These coordinates describe to an observer 
at  $\mathbf r_i$  how the backbone turns at vertex  $\mathbf r_{i+1}$ to continue towards the vertex $\mathbf r_{i+2}$. 
When we repeat this construction for all Frenet spheres in the case of C$\alpha$ backbones in PDB, we obtain the 
statistical distribution of ($\kappa, \tau$)  values that are highly concentrated inside the annulus ${\mathbb A}$ on the 
tangent plane that we show in figure \ref{fig3}b; the annulus $\mathbb A$ is located approximatively between latitude 
angle values $ 0.57 < \kappa < 1.82$ (radians). The color intensity characterizes the statistical occurrence of 
($\kappa, \tau$) values in PDB, decreasing from red and yellow to blue and white. The region that falls exterior to $\mathbb A$ is 
sterically limited,  while the interior is sterically allowed but with very few entries. The major secondary structure regions, 
including $\alpha$-helices, $\beta$-strands and left-handed $\alpha_L$-helices are identified in the Figure~\ref{fig3}b. 
For example, repeated ($\kappa,\tau$) values with
\begin{equation}
\left\{ \ \begin{matrix} \kappa_i  & \approx & \frac{\pi}{2} \\[1mm]
\tau_i & \approx & 1
\end{matrix} \right.
\label{alphahel}
\end{equation}
correspond to right-handed $\alpha$-helices, and repeated
values just below the branch cut  of the torsion angle, with
\begin{equation}
\left\{ \ \begin{matrix} \kappa_i  & \approx & 1 \\
\tau_i & \approx & \pm \pi
\end{matrix} \right.
\label{betas}
\end{equation}
correspond to $\beta$-strands that are stabilized through hydrogen bonding interactions. Similarly, all the 
other regular secondary structures, such as 3/10 helices, $\alpha_L$ helices {\it etc.} correspond to localized
 regions around fixed values of bond and torsion angles on $\mathbb A$. 
%
%
 
 \subsection{Local topology and perestroikas in discrete chains}

In the case of a regular space curve, the curvature and torsion are real-valued differentiable functions. The points 
where the local topology can change by perestroika are identified in terms of their zeroes or the zeroes in their derivatives. 
Similarly,  in the case of a discrete chain, we can have vertices and links where the local topology can change by appropriate 
perestroikas. For example, in line with differentiable curves, an inflection point occurs when two consecutive tangent vectors 
$\mathbf t_i$ and $\mathbf t_{i+1}$ become parallel, and in Figure \ref{fig3}b, this takes place at the center of the annulus 
$\mathbb A$ where $\kappa_i=0$. Similarly, a discrete variant of a flattening point occurs when two consecutive binormal 
vectors $\mathbf b_i$ and $\mathbf b_{i+1}$ become parallel so that $\tau_i=0$; the flattening line where this can take 
place is shown in Figure \ref{fig3}b. Moreover, unlike the Frenet torsion, which is a $\mathbb R^1$-valued function, the 
torsion angle is a multivalued variable taking values on $\mathbb S^1$, and we have a branch cut at $\tau=\pm \pi$ shown 
in Figure \ref{fig3}b. A branch point is then a point along the branch cut where two neighboring binormal vectors 
$\mathbf b_i$ and $\mathbf b_{i+1}$ become antiparallel. Notably, the values (\ref{betas}) of the  $\beta$-strand region are 
located in the vicinity of the branch cut. 

We may think 
of the small curvature and large torsion limit that we have described in subsection \ref{subsect:sectlimit} as a continuum
analog of the branch cut.  

In the sequel, whenever we relate the present formalism of local topology and ensuing perestroikas to actual experimentally 
measured protein structures, it is essential to keep in mind that in the case of observational data such as crystallographic 
protein structures, the positions of the C$\alpha$-atoms are only known with some experimental precision.  Thus, in the case 
of protein structures, a precise identification of an actual vertex where perestroika occurs may not be practical,  not even 
possible. Accordingly, to account for the experimental uncertainties, we consider any data point in Figure~\ref{fig3}b with a 
very small value of  $\kappa_i \approx 0$ to be a putative inflection point, any data point with $\tau_i \approx 0$ to be a 
putative flattening point and any data point with  $\tau_i \approx \pm \pi$ to be a putative branch point.  We observe from 
Figure \ref{fig3}b that PDB data in the immediate vicinity of an inflection point is absent, and data near the flattening line is rare,  
but data near the branch cut is relatively common, reflecting the abundance of $\beta$-strands.

\begin{figure}[h]
        \centering
                \includegraphics[width=0.48\textwidth]{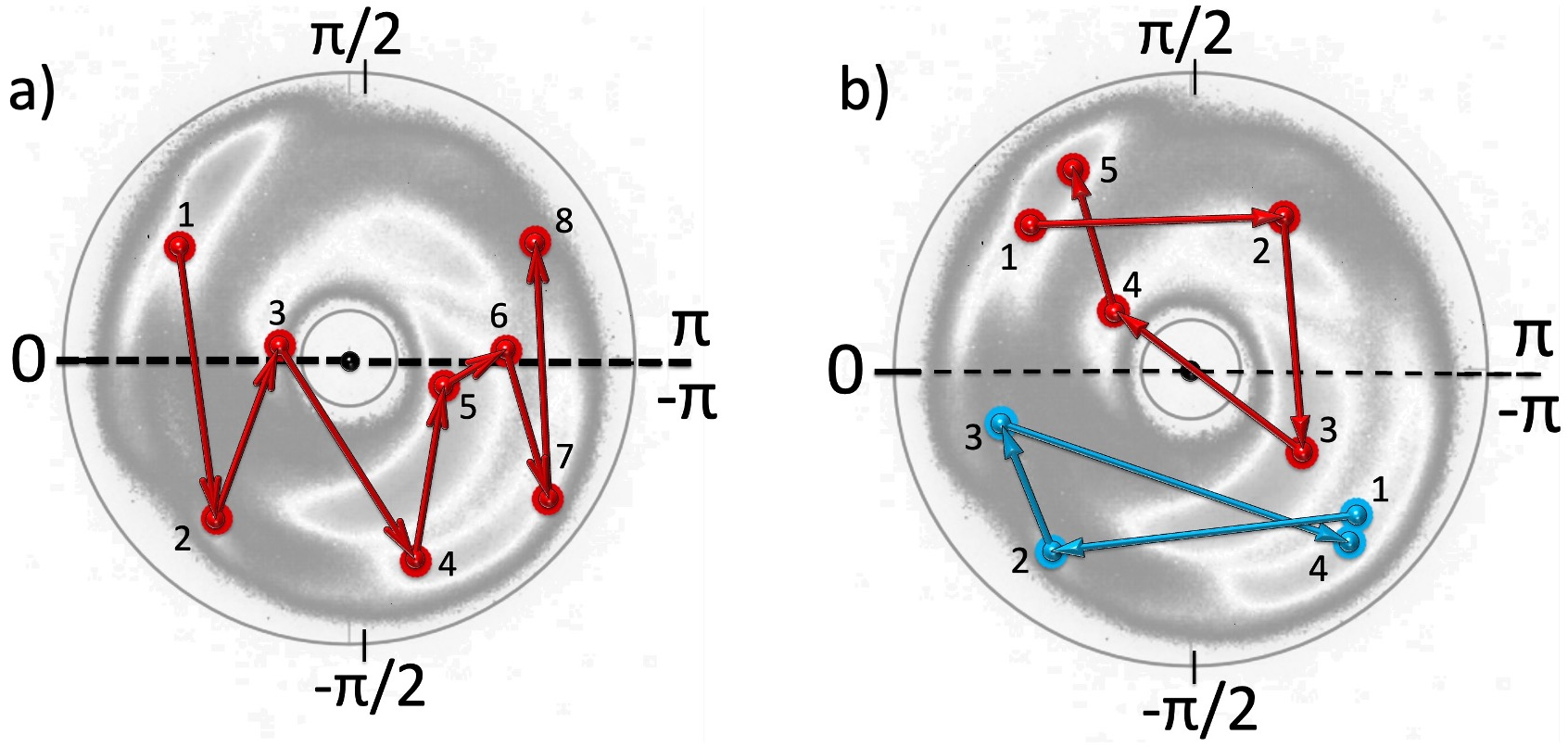}
        \caption{{
     Panel a: A generic example of a discrete C$\alpha$-chain depicted as a trajectory on the 
     annulus $\mathbb A$ of Figure \ref{fig3}a.
     Panel b: Two generic examples of a discrete C$\alpha$-chain depicted as trajectories on 
     the annulus. The red-colored chain starts and ends in the $\alpha$-helical region, the blue-colored
chain starts and ends in the left-handed $\alpha$-helical region.
}}
       \label{fig4}
\end{figure}

The Figures \ref{fig4}a and \ref{fig4}b exemplify short protein C$\alpha$ backbone segments, 
depicted as piecewise linear chains on the annulus $\mathbb A$  of Figure \ref{fig3}b;  the 
vertices are the ($\kappa_i,\tau_i$) values of the protein's C$\alpha$-atoms, that we connect 
by straight links on $\mathbb A$.

Note that a straight link between two C$\alpha$-vertices on $\mathbb A$  traverses over a range of bond and torsion 
angle values. It does not directly describe the diagonal line of the peptide plane that bonds the C$\alpha$-atoms in the 
physical $\mathbb R^3$ space. Nevertheless, we find the representation of a protein C$\alpha$ backbone in terms of a 
piecewise linear chain on the annulus $\mathbb A$, with vertices corresponding to the ($\kappa,\tau$) values of the 
C$\alpha$-atoms, to be very informative. It enables us to
conveniently extend Arnol'd's perestroikas to  piecewise  linear chains, as demonstrated by the  three 
examples shown in Figures~\ref{fig4}. 

\vskip 0.2cm
The Figure \ref{fig4}a depicts a short peptide chain on $\mathbb A$, starting at vertex 1 and ending at vertex 8. 
%
%
%
%
%
%
%
%
%
%
%
%
%
%
%
%
%
%
%
%
%
%
%
%
%
%
%
%

\vskip 0.2cm
$\bullet~$ The first link connects vertex 1 to vertex 2, proceeding across the flattening line $\tau=0$.  Clearly, this 
crossing across the flattening line persists when vertices 1 and 2 move around on $\mathbb A$, as long as there is 
no change in the sign of $\tau_{1}$ and $\tau_2$. Thus, the presence of a flattening 
point along a link that connects two vertices on the annulus $\mathbb A$ is a local topological invariant of the chain.

\vskip 0.2cm

$\bullet~$ The vertex 3 with small positive torsion angle $\tau_3$ is connected to vertices 2 and 4 by links that both 
cross the flattening line. Now, a small change in the position of vertex 3 can change the sign of its torsion angle if the 
vertex crosses the flattening line. If this occurs, we recognize the structure of a bi-flattening perestroika, which 
removes two flattening points 
along the chain.

\vskip 0.2cm

$\bullet~$ Next, we focus on vertices 4, 5, 6, and 7 in Figure~\ref{fig4}a.  There are now two branch points; one is 
along the link from vertex 5 to vertex 6, and the other is along the link between vertices 6 and 7. If the vertex 6 with 
$\tau_6>0$ initially crosses the branch cut so that its torsion angle becomes negative, we have a bi-branching 
perestroika akin to the bi-flattening perestroika where $\tau_3$ changes sign with vertex 3 moving across the 
flattening line. If, instead, the torsion angle at vertex 5 crosses the branch cut, and $\tau_5$ becomes positive, 
there is no perestroika of the chain; the two branch points simply move apart from each other along the chain.

Due to the ubiquity of $\beta$-strands, a conformation where two neighboring vertices are located in the immediate 
vicinity of the branch cut, even on opposite sides of it, like vertices 5 and 6 in Figure \ref{fig4}a, can be expected to 
be commonplace in the case of protein backbones.  

\vskip 0.2cm

$\bullet~$ The link connecting vertices 7 and 8 crosses the branch cut so that we have a branch 
point along the link. Like a flattening point between vertices 1 and 2, the presence of a branch 
point along the chain is a local topological invariant.  

\vskip 0.2cm

$\bullet~$ Notably, neither a bi-flattening point nor a bi-branching point 
is a local topological invariant: When the shape of the chain changes so that either a pair of 
flattening points or a pair of branch points come together, they combine either into a single 
bi-flattening point or a single bi-branching point. Both the bi-flattening point and the bi-branching 
point can then be removed by a further local deformation of the chain.

\vskip 0.2cm
In Figure \ref{fig4}b, we have two (essentially) closed chains on the annulus $\mathbb A$. This 
motivates us to introduce the concept of a {\it Folding Index} \cite{lundgren2013topology}. It is 
a local topological invariant 
that can be used to analyze and classify both chain segments and entire chains.   
To define this quantity, we consider a chain segment between two vertices, $n_1$ and $n_2$. 
Its folding index $Ind_f$ is then evaluated as follows, 
\begin{equation*}
Ind_f = \left[\frac{\Gamma(n_1,n_2) }{\pi}\right] 
\end{equation*}
\begin{equation}
\Gamma(n_1,n_2) \  = \ \sum\limits_{i=n_1}^{n_2-1} \begin{cases}
     \tau_{i+1}-\tau_{i}-2\pi & {\rm if}\   \tau_{i+1}-\tau_{i} > \pi\\
     \tau_{i+1}-\tau_{i}+2\pi & {\rm if}\  \tau_{i+1}-\tau_{i} < -\pi\\
     \tau_{i+1}-\tau_{i} & \ \ \ \ {\rm otherwise}
\end{cases}
\label{foldind}
\end{equation}
Here  $[x]$  denotes the integer part of $x$, and $\Gamma(n_1,n_2)$  is the total rotation angle (in 
radians) that the chain segment winds around the inflection point {\it i.e.} the center of the annulus as it 
proceeds from vertex $n_1$ to vertex $n_2$. A clockwise winding is positive, and winding in the 
counterclockwise direction is negative, and the value of the folding index is equal to twice the net 
number of times a chain encircles the center of the annulus $\mathbb A$.

\vskip 0.2cm
$\bullet~$  In Figure \ref{fig4}b,  we first consider the red-colored chain starting from vertex 1 and 
ending in vertex 5,  both in the $\alpha$-helical region. The chain depicts a protein loop structure from vertex 1
through vertex 2 to vertex 3 that is located in the  $\beta$-stranded region, representing an 
$\alpha$-helix - loop - $\beta$-strand supersecondary structure. The link between vertices 2 and 3 
proceeds through the branch cut, and the link between vertices 3 and 4 passes through the center of the 
annulus, {\it i.e.} there is an inflection point between vertices 3 and 4. Note that a slight move of either 
vertex 3 or vertex 4 horizontally removes this inflection point, and we are left either with a flattening point or 
a branch point along the link between vertices 3 and 4. In the first case, the folding index of the segment would 
obtain the value +2, while in the second case, the folding index would vanish. 

In general, if the torsion angles at the endpoints of a one-parameter family of chains are kept fixed, 
the value of the folding index can only change at an inflection point perestroika, {\it i.e.} when the chain 
is deformed so that a link passes over the center of the annulus $\mathbb A$. When this occurs, the 
folding index changes by $\pm 2$, and a flattening point becomes converted into a branch point or {\it vice 
versa}. In particular, an inflection point is not a local topological invariant.

\vskip 0.2cm
$\bullet~$ The blue chain in Figure \ref{fig4}b 
shows an example of a loop structure starting and ending in the left-handed $\alpha$-helical region.  
The chain closes, but since it does not cross the flattening line nor the branch cut, and in particular, it 
does not wind around the inflection point, its folding index vanishes.

\vskip .2cm

%
 
 \section{Local topology of myoglobin}
 
 \subsection{Proteins and fractals}

Topology characterizes the overall shape of a protein C$\alpha$ backbone, usually in a manner 
that is quite independent of the individual atom positions that govern the backbone geometry. For 
this reason, it is possible to describe general topological aspects of globular proteins using a generic 
example, and we chose the crystallographic myoglobin structure with Protein Data Bank accession 
code 1ABS. It comes from a sperm whale, has 154 amino acids,  and has been measured at about 
20 Kelvin. This ultra-low temperature setting significantly minimizes the experimental errors in the 
measured coordinates of individual atoms, ensuring high precision at the level of the geometric data. 
Our description of 1ABS builds partly upon previous theoretical investigations of its structure in 
\cite{Krokhotin-2013,Begun-2019}. 
%

%
%
%
%
%
%
%
%
%
%
%
%
%
%
%
%
%
%
%
%
%
%
%
\begin{figure}[h]
        \centering
                \includegraphics[width=0.45\textwidth]{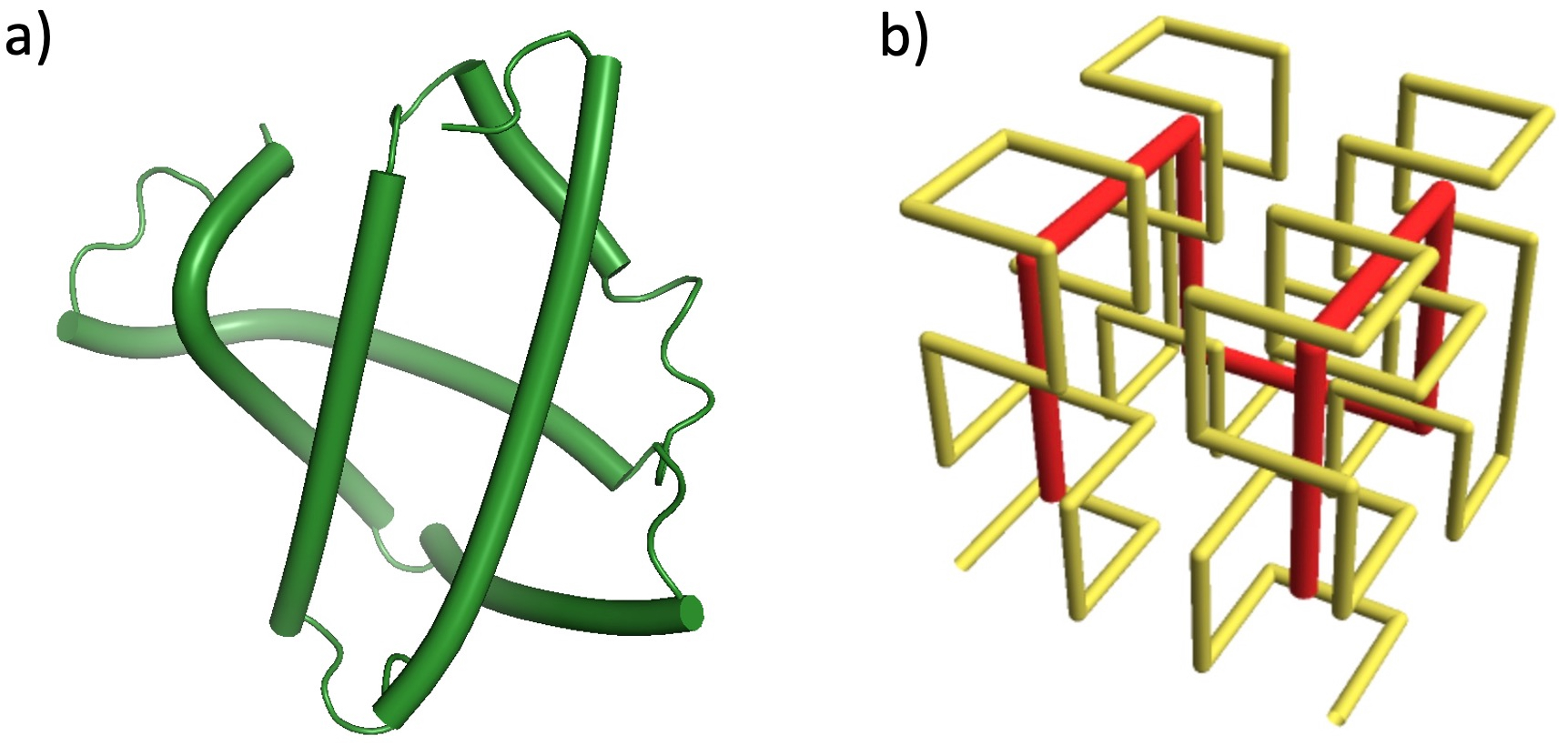}
        \caption{{
    a) Cartoon representation of 1ABS C$\alpha$ backbone, with repetitive helical segments 
    replaced by linear segments for visual clarity.
     b) First (red) and second (yellow) iterations of the space-filling Hilbert curve.
 }}
       \label{fig5}
\end{figure}
%
%
%
%
%
Figure~\ref{fig5}a shows the 1ABS C$\alpha$ backbone in a cartoon representation where we 
have removed the atoms along the helices in order to visually enhance the backbone's repetitive 
propagation through its regular secondary structures.   
The Figure identifies a key feature of globular proteins: They are commonly made up of building blocks 
called super-secondary motifs, consisting of regular secondary structures such as 
$\alpha$-helices and $\beta$-strands together with their
interconnecting turns and loops that appear irregular.  The fact that globular proteins are often built in a modular
fashion is supported by the success of protein structure classification schemes like CATH \cite{CATH} and 
SCOP \cite{SCOP}. The modularity can also be considered as one of the  explanations why artificial intelligence 
programs such as  
AlphaFold \cite{alpha} can predict folded protein structures.

The modularity of globular protein C$\alpha$ backbones in their cartoon representation is in a striking 
resemblance to the modular topology of space-filling Peano curves \cite{Sagan-2012}. We illustrate 
this in Figure \ref{fig5}b, where we display the first (red) and second (yellow) iteration steps of the Hilbert 
curve. The first step is a combination of alternating linear segments connected by turns, akin to a protein 
super-secondary motif,  and the second step is a self-similar iteration of the first step together with an 
overall scaling transformation. In the limit of an infinite number of iterations, a Peano curve becomes 
a space-filling chain with Hausdorff dimension $D_H=3$,  and several analyses based on
investigations of the radius of gyration (\ref{Rg})  propose that 
folded proteins share this  Hausdorff dimension \cite{Huang}.  The apparent similarity between globular 
proteins and Peano curves, with a topology of linear segments that are connected by loops and turns, leads us to propose 
that the modular framework of  Peano curves is the appropriate setting for understanding the rationale
 in the modularity of protein C$\alpha$ backbone structures, explaining their space-filling character.

%
%
%
%
%
%
%
%
%
%
%
%
%
%
%
%
%
%
%
%
%
%
%
\begin{figure}[h]
        \centering
                \includegraphics[width=0.45\textwidth]{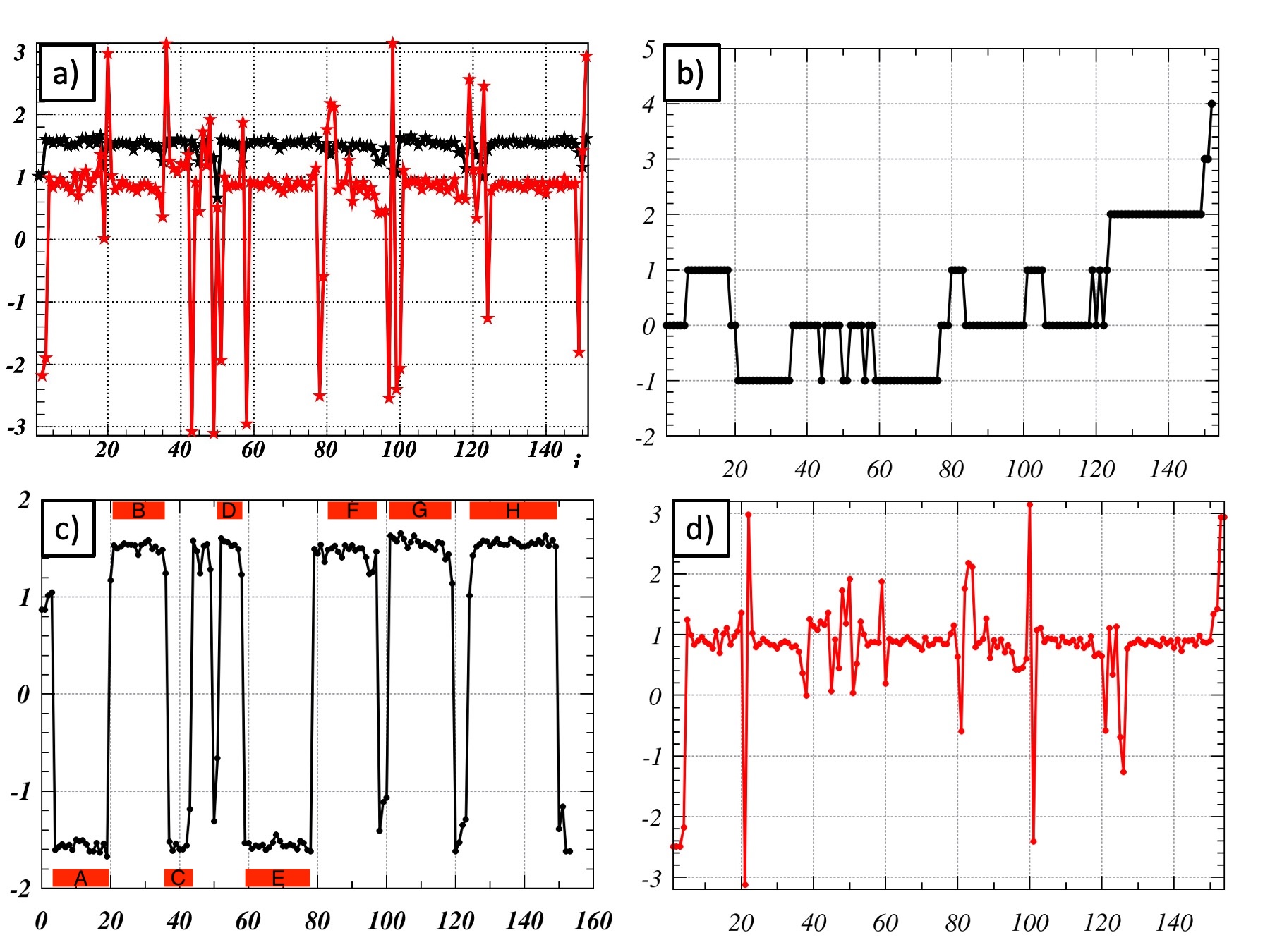}
        \caption{{
     a) Discrete Frenet frame bond angles (black) and torsion angles (red) along the 1ABS C$\alpha$ backbone  b) 
     Folding index along 1ABS C$\alpha$ backbone c) $\mathbb Z_2$ transformed bond angles and d) 
      $\mathbb Z_2$ transformed torsion angles, with  $\mathbb Z_2$ transformations determined by minimization of (\ref{tordis}). 
      The panel c) also identifies the common assignment of the eight helices A, \dots , H along the backbone.            }}
       \label{fig6}
\end{figure}

\subsection{The $\mathbb Z_2$ transformation}

To quantify the modular building blocks of globular proteins, 
in Figure \ref{fig6}a, we present the spectrum of the discrete Frenet frame bond angles (\ref{bond}) and torsion angles (\ref{tors}) of 
the 1ABS C$\alpha$ backbone. This figure corroborates the well-established observation  
that  the bond angles in protein C$\alpha$ backbones exhibit significantly more rigidity than the torsion 
angles. 
%

In Figure \ref{fig6}b we show the accumulation 
of the  folding index along the 1ABS backbone, starting from the 
N-terminus. The final value is $Ind_f  = +4$  so that the entire backbone trajectory encircles 
twice the center of the annulus $\mathbb A$  in Figure \ref{fig3}b
in the clockwise direction. A segment where the folding index along the chain does not change is the hallmark 
of a regular secondary structure such as $\alpha$-helix and $\beta$-strand, while the segments where the 
folding index changes identify loop structures. Notably, there are four segments in
Figure \ref{fig6}b where we observe the change $\Delta Ind_f = \pm2$,  implying that there is a complete 
encirclement of the center of the annulus $\mathbb A$; one of these occurs at the C-terminal.

Next, we consider the consequences of the $\mathbb Z_2$ transformation (\ref{dsgau}); recall that this transformation is a 
symmetry of the discrete Frenet equation (\ref{DFE2}) that does not affect the three-dimensional shape of the chain. 
At the same time, it turns out to be a key to understanding the origin of modularity in protein structure and its 
connection to the local topology of the backbone. 
By extending the bond angles to negative values, the $\mathbb Z_2$ transformation also entails an extension of the 
annulus  $\mathbb A$ into its two-sheeted covering space.

As a methodology to select the way how to implement the
$\mathbb Z_2$ transformation, we consider the following metric that measures 
distances in the space of torsion angles between a vertex $n_1$ and a vertex $n_2$
(with $|\tau_{i+1}-\tau_{i}| \leq 2\pi$)
\[
\delta_\tau (n_1,n_2)  = \sum\limits_{i=n_1}^{n_2-1}\delta_{\tau,i}  
\]
where
\begin{equation}
\delta_{\tau,i} = 
\begin{cases}
   \  |\tau_{i+1}-\tau_{i}-2\pi|^2\ \ & \ \ {\rm if}\   \tau_{i+1}-\tau_{i} > \pi \\
   \  |\tau_{i+1}-\tau_{i}+2\pi|^2 \ \ & \ \ {\rm if}\   \tau_{i+1}-\tau_{i} < -\pi\\
   \  |\tau_{i+1}-\tau_{i}|^2 & \ \ \ \ \ \ \ \  {\rm otherwise}
\end{cases}
\label{tordis}
\end{equation}
Unlike the folding index that does not admit any practical analog in the continuum limit, as there is no branch cut, 
the continuum limit of  the metric (\ref{tordis}) corresponds to the elastic torsion energy 
\[
E_\tau \ = \ \int\limits_{s_1}^{s_2} ds \,  (\partial_s \vartheta)^2
\]

Since the metric (\ref{tordis}) does not remain invariant under the $\mathbb Z_2$ transformation, we can consider 
the minimization of  $\delta_\tau$ over all possible $\mathbb Z_2$ transformations. In  Figure \ref{fig6}c and \ref{fig6}d, 
we show the results of the minimization for the bond and torsion angles, respectively, in the case of 1ABS. At the 
minimum of $\delta_\tau$, there are a total of 13 bond angle domain walls that separate regions with positive and 
negative $\kappa_i$ values when we include the two that are right next to the N and C terminals. As expected, the 
profile of the $\mathbb Z_2$ transformed bond angles correlates with the folding index profile in Figure \ref{fig6}b.

Structural biology textbooks commonly identify eight helices (A, B, C, D, E, F, G, H) along the myoglobin backbone. 
These helices are interconnected by seven loops, and in addition, there are the unstructured  N and C terminals. 
The eight standard helices are identified in Figure \ref{fig6}c, and they also appear as helices in our approach. In 
the textbook assignment of helices and loops, the short segment between helices C and D is interpreted as a 
single loop.  But in our refined approach, based on minimization of (\ref{tordis}),  this loop becomes resolved into 
three adjoining domain walls. Furthermore, the two loops between helices F and G, and G and H both become 
resolved into combinations of two domain walls each in our approach. 

More generally,  in the case of globular proteins, long loops become commonly resolved into combinations of 
multiple individual domain walls when we minimize the torsion distance~(\ref{tordis}).

\subsection{Free energy for local topology}

All-atom molecular dynamics simulation techniques, such as GROMACS \cite{gromacs}, 
could be used to explore the role of local topology and perestroikas in protein dynamics and folding processes. 
These techniques are designed to capture the intricate geometrical details of protein structure and dynamics at 
the level of individual atoms, and for this, they require significant computational resources that are not widely available. 
Therefore, a more efficient approach is to adopt a physics-based model that focuses on the structure and dynamics 
of a  protein C$\alpha$ backbone, directly employing variables that are pivotal for understanding its local topology. 
Furthermore, since topology is hardly sensitive to an individual atom precision, this is the approach that we adopt.

The $\mathbb{Z}_2$-transformed $\kappa_i$ domain walls, depicted in 
Figure \ref{fig6}c, exhibit a profile reminiscent of a discretized topological soliton akin to that shown in 
Figure \ref{fig1}a. This proposes to model the C$\alpha$ backbone as a multi-soliton system using a suitably defined 
discrete version of the free energy (\ref{Fkink}).  Indeed, such a framework has been introduced in references 
\cite{Danielsson-2009,Molkenthin-2011}, and its applicability in modeling the 1ABS C$\alpha$ backbone has been 
demonstrated in \cite{Krokhotin-2013,Peng-2016,Begun-2019}. Moreover, in combination with stereochemical 
considerations \cite{Hou-2019}, the approach can reach a sub-\AA ngstr\"om precision even in the case of all-atom 
structures, matching the accuracy that can be obtained by direct minimization of the potential energy in all-atom force fields.   

The starting point in the construction of the free energy~\cite{Danielsson-2009} is the observation that a complex 
variable such as (\ref{phivar}) also appears as the dynamical variable in nonlinear Schr\"odinger equation \cite{Manton}, 
supporting a dark soliton solution with the hyperbolic tangent profile (\ref{kink2}). This motivated the introduction of the 
following discretized nonlinear Schr\"odinger (DNLS) free energy for protein C$\alpha$ backbones
\cite{Danielsson-2009}
\begin{align}
F(\kappa,\tau) = & - \sum\limits_{i=1}^{N-1}  2 \kappa_{i+1} \kappa_{i} + 
\sum\limits_{i=1}^N \, { \bigg \{  }  2\kappa_i^2 + 
 \lambda\, (\kappa_{i}^2 - m^2)^2  \nonumber\\
 & + \frac{d}{2} \, \kappa_{i}^2 \tau_{i}^2   
- b \, \kappa^2_i \tau_{i}   - a \,  \tau_{i} +  \frac{c}{2}  \tau^2_{i} 
 \bigg \}\,. 
\label{Esol}
\end{align}
The numerical values of the parameters ($\lambda,m,d,b,a,c$) in (\ref{Esol}), that are specific for each individual
soliton profile,  
are determined by  
demanding that the minimum of  (\ref{Esol}) describes the given protein C$\alpha$ backbone,
in our case 1ABS,  as a critical point
with the desired precision \cite{Molkenthin-2011}; less than 1.0\AA~root-mean-square-distance (RMSD) in the case of 1ABS.

To solve for the appropriate minimal energy critical point of (\ref{Esol}), we note that the bond angles 
are relatively rigid and slowly varying while 
the torsion angles are quite flexible and
with rapid variations. Thus, we may construct the free energy minimizer in the adiabatic approximation,
by first eliminating the torsion angles in terms of the bond angles,
\begin{equation}
\frac{\partial F}{\partial \tau_i} = d\kappa_i^2 \tau_i - b\kappa_i^2 - a + c\tau_i = 0
\label{dfdt}
\end{equation}
from which
\begin{equation}
\tau_i[\kappa] = \frac{ a+b\kappa_i^2}{c+d\kappa_i^2}
\label{tausol}
\end{equation}
We substitute this in (\ref{Esol}) and find for the bond angles the following free energy
\begin{equation} 
F[\kappa] = - \sum\limits_{i=1}^{N-1}  2 \kappa_{i+1} \kappa_{i} + 
\sum\limits_{i=1}^N \, \left( 2\kappa_i^2 +  V[\kappa_i]\right)\,,
\label{Fk}
\end{equation}
where 
\begin{equation}
V[\kappa] = \lambda \kappa^4 - \left(
 \frac{ b^2+8\lambda m^2} {2b} \right) \, \kappa^2 - \left( \frac{ bc-ad}{d} \right)  \frac{1}{c+d\kappa^2}\,.
\label{VKpot}
\end{equation}
The minimum energy critical point of (\ref{Fk}) is a solution of the equation 
\begin{equation}
\kappa_{i+1} = 2\kappa_i - \kappa_{i-1} + \frac{ dV[\kappa]}{d\kappa_i^2} \kappa_i \ \ \ \ \ \ (i=1,\dots ,N)\,.
\label{dnls}
\end{equation}
with $\kappa_0=\kappa_{N+1}=0$. 
We identify here a  discrete variant of  (\ref{kink1}),  with an additional 
contribution due to torsion variables; 
in the case of C$\alpha$ backbones  the torsion contribution in (\ref{VKpot}) commonly
has a small numerical value in comparison to the first two terms in 
(\ref{VKpot}).  As a consequence 
the appropriate solution of (\ref{dnls}), while not known in analytic form, is 
very close to a combination of the topological soliton profiles 
(\ref{kink2}) along the C$\alpha$ backbone.

After solving (\ref{dnls}) the corresponding torsion angle values are
evaluated from (\ref{tausol}) and the space coordinates of the C$\alpha$ backbone are then 
computed using the discrete Frenet equation (\ref{DFE2}), (\ref{chain}). In \cite{Molkenthin-2011} it 
has been explained in detail 
how the multi-soliton and the ensuing C$\alpha$ backbone can be numerically constructed.  The 
individual solitons describe the super-secondary motifs, in the case of  generalized 
Peano curves  they would correspond to the modular building blocks. 

\subsection{Myoglobin as a multi-soliton}

As shown in Figure \ref{fig6}c in the case of 1ABS the minimization of (\ref{tordis}) results in 11 solitons, 
excluding those at the flexible termini, with the loop between C and D segmented into three individual solitons. 
But the 10-soliton energy minimum of (\ref{Esol}) constructed in \cite{Krokhotin-2013, Peng-2016, Begun-2019} that
combines the soliton pair around site 45 into a single soliton, 
already represents the 1ABS backbone with a RMSD of 0.8 Å ~ {\it i.e.} well below the experimental 
resolution of 1.5 Å.  Since these 10 solitons suffice for accuracy that surpasses experimental resolution, there is no 
need to introduce an additional one, and in line with  \cite{Krokhotin-2013, Peng-2016, Begun-2019}, we adopt the 10-soliton 
representation of 1ABS.

\begin{figure}[h]
        \centering
                \includegraphics[width=0.45\textwidth]{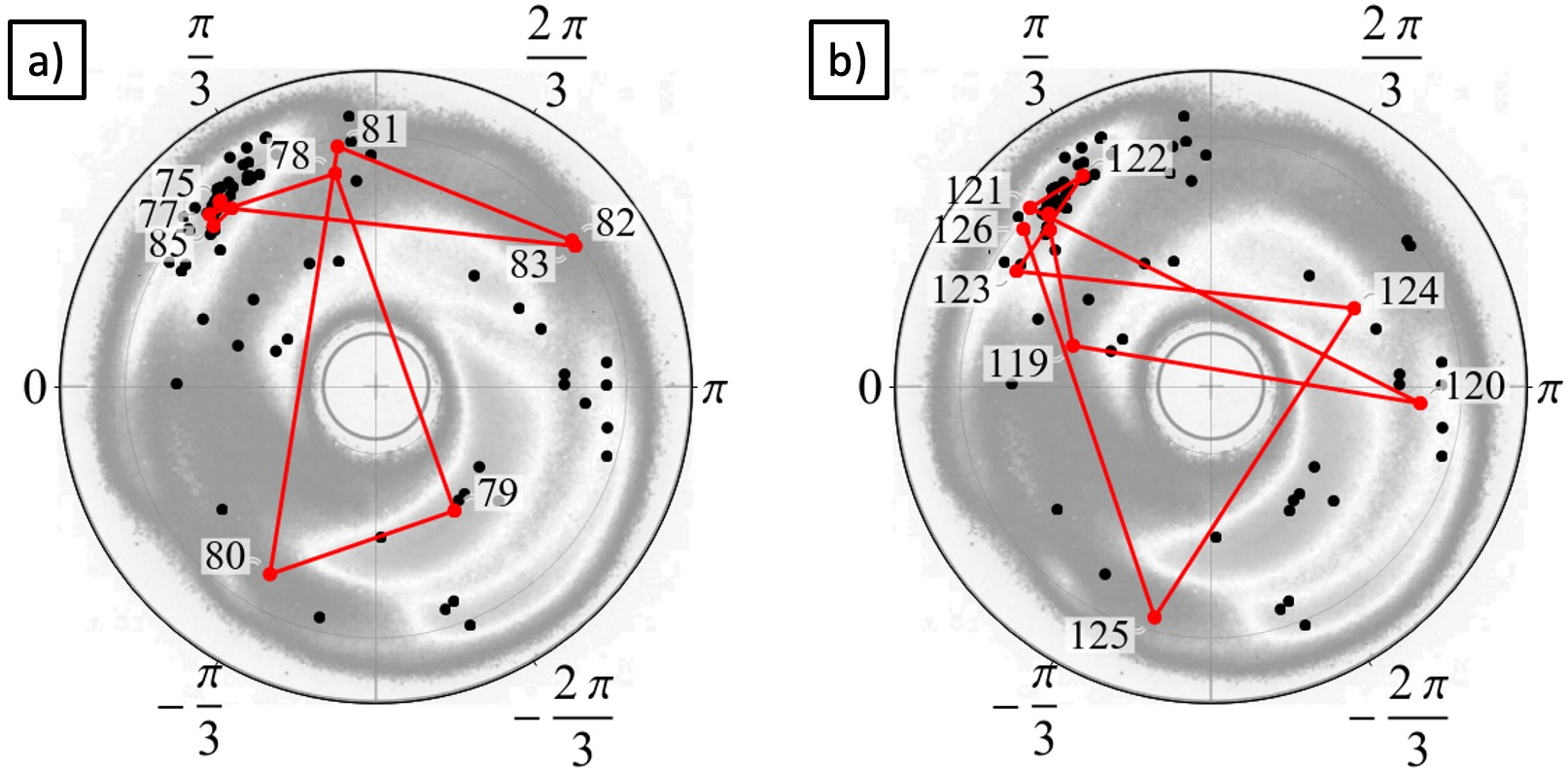}
        \caption{{
     a) The segment 75-85 of 1ABS, shown as a trajectory on the annulus $\mathbb A$. b) The segment 115-126 of 1ABS. 
     The black dots denote the positions of the remaining C$\alpha$ atoms on $\mathbb A$. 
            }}
       \label{fig7}
\end{figure}

To exemplify how a soliton describes the local topology of the C$\alpha$ backbone, we 
proceed to describe in detail the soliton that models 
the segments 75-85 between helices E and F, and the two-soliton that models the segments 115-126 between G 
and H. In Figures \ref{fig7}a and b, we show these segments as trajectories on the annulus $\mathbb A$.

%
%
%
%
%
%
%
%
%
%
%
%
%
%
%
%
%
%
%
%
%
%
%

The single soliton profile that models the segment 75-85 between helices E and F
is centered near site $i=80$, and as seen in Figure \ref{fig6}a
the folding index for the soliton segment has the value $Ind_f=+2$. 
The Figure \ref{fig7}a shows that the trajectory encircles the center of the annulus with a branch point along the 
link connecting vertices 78 and 79 and a flattening point along the link between sites 80 and 81. After site 81, there 
is also a short loop segment with two entries with $\tau \approx 3\pi/4$ prior to the trajectory proceeding to helix F 
in the $\alpha$-helical region.

The segment 115-126 between helices G and H consists of a soliton pair, one centered near site $i=121$ 
and the other near site $i=125$. Figure \ref{fig6}b shows that the folding index of the entire segment has the 
value $Ind_f=+2$. Indeed, as seen in Figure \ref{fig7}b, between vertices 123-126, the trajectory encircles
once the center of the annulus in the clockwise direction. The first soliton starts in the $\alpha$-helical 
region (helix G), and proceeds to $\beta$-stranded region. There is a branch point both along link 119-120 and 
along link 120-121; topologically, the segment between vertices 119 and 121 is similar to the segment between 
vertices 6-7 in Figure \ref{fig4}a. Furthermore, the link between vertices 119 and 120 passes very close to an inflection 
point. Thus, with $\tau_{120}$ persisting negative value, a small decrease in the $\tau_{119}$ value can bring 
about an inflection point perestroika that changes the folding index of the segment 115-126 
into $Ind_f=0$ by conversion of a branch point into a flattening point along the link between $i=119$ and $i=120$. 
This flattening point then moves to the link between $i=118$ and $i=119$, provided the value of $\tau_{119}$ 
decreases further so that it becomes negative.  Similarly, with the vertex 119 intact, a small increase in the 
$\tau_{120}$ value can also bring about an inflection point perestroika that changes the folding index of the 
segment into $Ind_f=0$ by converting the branch point into a flattening point on the link between vertices $i=119$ and $i=120$.

%
%
%
%
%
%
%

\section{Thermal dynamics and local topology}

\subsection{Glauber dynamics} 

Using the ambient temperature as a control parameter, 
we investigate changes in the local topology of the 1ABS backbone during thermal folding and 
unfolding; the ambient temperature then has the same 
role as the parameter $m$ in the example (\ref{saddle})-(\ref{Fkink}). We simulate thermal effects
using the Glauber algorithm \cite{Glauber-1963}. It models pure relaxation dynamics so that a 
C$\alpha$ backbone evolving according to the Glauber algorithm approaches the instantaneous Gibbsian 
thermal equilibrium state at an exponential rate. We realize Glauber dynamics using  a Monte Carlo 
algorithm that evaluates the transition probability from a conformational state $a$ to a conformational 
state $b$ using the following probability density~\cite{Berg-2004},
\begin{equation}
\mathcal P(a\to b) \ = \ \frac{1}{1+ e^{F_{ba}/T} }\,.
\label{glauber}
\end{equation}
The parameter $T$ in (\ref{glauber}) is the Monte Carlo temperature factor that acts as our control 
parameter. The activation energy $F_{ba}$ in (\ref{glauber}) is the difference between the corresponding 
free energies (\ref{Esol}) between the two states, augmented  as follows:
\begin{equation}
    F(\kappa , \tau ) \to F(\kappa,\tau | \mathbf r)
    \ = \ F(\kappa,\tau) + \sum\limits_{|i-j|\geq 2} V(\mathbf r_i - \mathbf r_j) \,.
    \label{Esol2}
\end{equation}
The two-body potential $V(\mathbf r_i - \mathbf r_j)$ can include various contributions, from the 
short-distance excluded volume interaction to the long-range Coulomb interaction, and we refer to 
\cite{Sinelnikova-2018} for a discussion on different 
two-body potentials in the present context. In our simulations, with  focus on local topology 
and not on geometric atomic level details, it is sufficient to account for only the excluded volume 
repulsion that we model using the hard-core potential
\begin{equation}
   V(\mathbf r_i - \mathbf r_j) \ = \ \left\{
  \begin{matrix}   \left. \begin{matrix}
       \infty \hspace{0.55cm} {\rm if} \ \ |\mathbf r_i - \mathbf r_j | \leq \Delta \\
        \, 0 \hspace{0.65cm} {\rm if} \ \ |\mathbf r_i - \mathbf r_j| > \Delta \end{matrix}
    \ \ \ \ \ 
     \right. 
     |i-j| \geq 2 \\  
       \hspace{0.13cm} 0 \ \ \ \ \, \hspace{3.3cm}   |i-j|<2 \end{matrix} \right.
\label{Vr}
\end{equation}
and we choose $\Delta = 3.7$\AA ~ that coincides with the diagonal length of the peptide plane, 
ensuring that the distance between any two C$\alpha$ atoms that are not nearest neighbours along 
the backbone can never be less than the distance between two neighboring C$\alpha$ atoms. The 
units in (\ref{glauber}) are set by the overall scale of (\ref{Esol}), and we refer to \cite{Peng-2016} for 
a detailed relation between $T$ and the physical temperature factor $k_B T_K$ with $k_B$ the 
Boltzmann constant and $T_K$ the ambient temperature measured in Kelvin. 

\subsection{Geometric order parameters} 

\begin{figure}[h]
        \centering
                \includegraphics[width=0.48\textwidth]{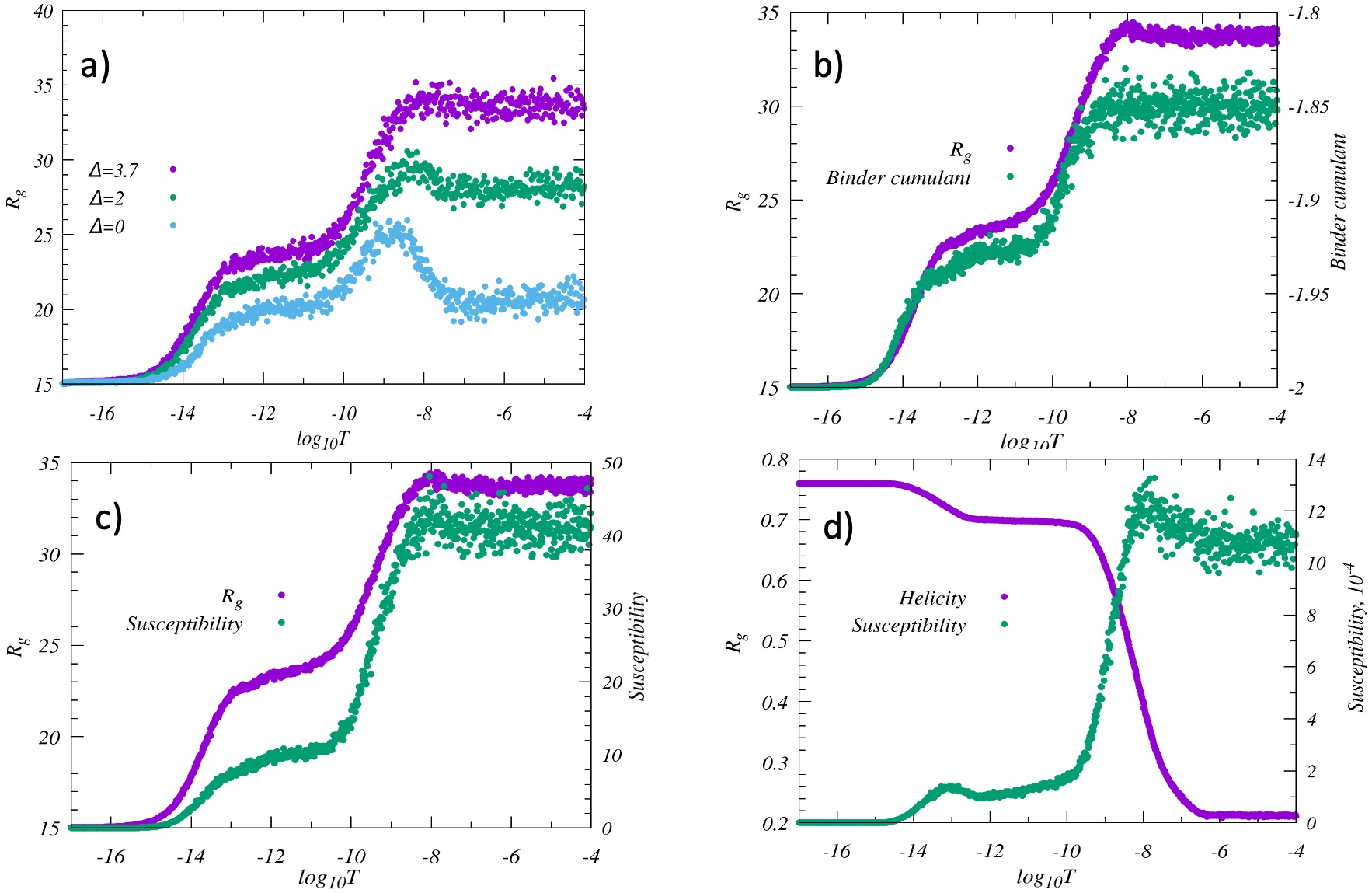}
%
%
%
%
%
%
%
%
%
%
%
%
%
%
        \caption{{
     a) The radius of gyration $R_g$ as a function of the Monte Carlo temperature factor $T$ for  $\Delta = 3.7$(\AA)
     that is used in our 1ABS simulations, $\Delta = 2.0$(\AA) as a generic example of a small $\Delta$, 
     and $\Delta = 0$ that approaches  random walk in the high-$T$ limit. b) The Binder cumulant of $R_g$ with 
     $\Delta=3.7$ as a function of $T$. c) The susceptibility of $R_g$ with $\Delta=3.7$ as a function of $T$. d) 
     The helicity together with its susceptibility as a function of $T$,  with $\Delta=3.7$.
            }}
       \label{fig8}
\end{figure}

First, we analyse  thermal effects on the protein backbone 
in terms conventional, geometry based order parameters {\it a.k.a.} reaction coordinates. 
We start with the radius of gyration (\ref{Rg}).  The Figure \ref{fig8}a shows its evolution
as a function of the temperature factor $T$ with three
different values of $\Delta$ in (\ref{Vr}).
%
%
%
%
%
%
%
%
%
%
%
%
%
%
%
%
%
%
%
%
%
%
%
%
%
%
%
%
%
%
%
%
%
%
%
%
The results for the  1ABS simulations {\it i.e.} with $\Delta = 3.7$(\AA)  are shown using purple data points. 
At low $T$ values, when  $\log_{10}T < -14$ we have $R_g \approx 15 $(\AA). This coincides with the radius
 of gyration of the crystallographic 1ABS in the folded, space-filling (Peano) phase; see (\ref{nuval}). 
 The low-temperature phase is followed by a transition regime around values $\log_{10} T \in [-15,-13]$, 
 where the radius of gyration increases to $R_{g} \sim 24$(\AA). This interval of temperatures corresponds to 
 the experimentally measured $R_g$ in the molten globule phase (\ref{nuval})~\cite{Peng-2016}.
The molten globule phase persists until the system enters a second transition regime with $\log_{10} T 
\in [-10,-8]$ where the radius of gyration increases to $R_g \sim 34$. We observe no change in $R_g$ 
when $T$ increases further; note that in the high-$T$ limit, only the excluded volume condition (\ref{Vr}) 
persists. Thus, in the large-$T$ regime, the backbone is in the phase of self-avoiding random walk (SARW); see
(\ref{nuval}).

The simulation results for the value $\Delta = 2.0$(\AA) are presented by the green data points in 
Figure (\ref{fig8})a. This value of $\Delta$ is chosen to exemplify the effects of the self-avoiding 
condition in terms of a generic non-vanishing but small $\Delta$ value. However, it is important to 
keep in mind that in the case of an 
actual protein backbone, any distance less than $\Delta = 3.7$(\AA) between any two C$\alpha$ atoms 
along the backbone chain is normally sterically excluded \cite{Hou-2019}. For low temperatures, up 
until $\log_{10} T \approx -9$, the $R_g$  values are quite similar to those with $\Delta = 3.7$(\AA); 
the difference can be interpreted as an effective $\Delta$-dependence of the Kuhn length $R_0$ in 
(\ref{R}). When $T$ increases to $\log_{10}T \in [-9,-8]$ the value of $R_g$ decreases slightly and 
then stabilizes to 
$R_g \approx 28$(\AA)  which is the high temperature SARW value for $\Delta = 2.0$(\AA).

For comparison, the light-blue data in Figure (\ref{fig8})a describes the case $\Delta =0$, with no 
self-avoidance. Up to $\log_{10}T \approx -9$ the values of $R_g$ evolve in line to $\Delta = 3.7$ and 
$\Delta=2.0$ reaching  $R_g \approx 25$(\AA); the difference to $\Delta = 3.7$ and $2.0$ can again be 
interpreted in terms of an effective $\Delta$-dependency of the Kuhn length (\ref{R}). But 
after $\log_{10}T > -9$ the value of $R_g$ starts decreasing until $\log_{10} T \approx -7$, after which it 
stabilizes to $R_g \approx 20$(\AA). Since neither excluded volume repulsion nor the free energy 
contributes in the large-$T$ limit, with $\Delta=0$, the large-$T$ regime describes a chain that resides 
in the fully flexible, Brownian  random  walk phase.

Remarkably, in the case of $\Delta=0$, even though the radius of gyration values in the 
intermediate $\log_{10}T\in [-12.5, -10]$ regime and in the large-$T$ regime are the same, these two 
regimes clearly represent two different phases that are separated by a transition region of variable 
$R_g$, peaked at around $\log_{10}T \approx -9$. Since the $\log_{10}T \in [-12.5, -10]$ regime of 
$\Delta=0$ connects smoothly to the molten globule phase of 1ABS, as shown by the $\Delta=2$ 
intermediate, we also conclude that in the case of myoglobin, the molten globule phase is different from 
the random walk phase, even if the two share the same  scaling exponent $\nu$ of (\ref{Rg}): The molten globule phase 
describes a self-avoiding walk with scaling exponent  $\nu\approx 1/2$. 

In summary, from the results of  Figure \ref{fig8}a, we deduce that myoglobin ($\Delta=3.7$) has three 
geometrically distinct phases that are separated by two intermediate transition regimes as follows: 
There is the low-$T$ folded phase. There is an intermediate-$T$ molten globule phase where the radius 
of gyration has the same value as in the random walk phase, but the two phases are different. Finally, 
there is the high-temperature self-avoiding random walk phase. 

We have also investigated the phase structure using the lowest order Binder cumulant $B_Q$~\cite{Binder-1981}. 
In the case of a quantity $Q$, this Binder cumulant  is defined as follows:
\begin{equation}
B_Q = \frac{\langle Q^4\rangle }{\langle Q ^2\rangle ^2} - 3\,,
\label{BQ}
\end{equation}
The Binder cumulants are commonly employed to identify different phases and phase transitions in statistical 
systems. These cumulants are especially useful in the case of finite systems, such as a protein in our case, as they  can often 
identify phase transitions and describe critical phenomena without a need for an extensive finite-size scaling 
analysis.  With the normalization of Eq.~\eqref{BQ}, in the case of an ordered phase such as the ferromagnetic 
phase in a magnetic system, this Binder cumulant tends towards $B_Q = -2$ while in a disordered phase it
approaches zero; a higher value of $B_Q$ generally indicates a higher degree of disorder.  

In Figure \ref{fig8}b, we show the $T$-evolution of the Binder cumulant $B_R$ evaluated for the radius of 
gyration $R_g$, in the case of myoglobin. The figure confirms the existence of the three distinct phases 
that we have identified in Figure \ref{fig8}a,  as the three (approximate) constant valued levels of $R_g$. 
In the collapsed low-temperature phase, the Binder cumulant has the value $B_R=- 2$ confirming that 
this is an ordered phase. In the intermediate molten globule phase, the value of Binder cumulant is 
$B_R\approx -1.93$, and in the high temperature self-avoiding random walk phase $B_R \approx -1.85$. 
Notably, in these two apparently disordered phases we still have relatively low Binder cumulant values so 
that there is a degree of order present. 

Besides the radius of gyration, there are also other geometric order parameters that can be introduced, 
and as an example we consider the helicity $H$ that we define as follows: A C$\alpha$ atom is in a helical 
position if the value of its bond angle is within the range $\kappa\in [1.29, 1.78]$, and the value of its torsion 
angle is in the range $\tau \in [0.5, 1.27]$. These values are selected to cover the $\alpha$-helical region 
on the annulus $\mathbb A$ in Figure \ref{fig3}b; other values could also be considered. The helicity $H$ is 
then the ratio in the number of those C$\alpha$ atoms that have ($\kappa,\tau$) values in the above region 
over the number of all C$\alpha$ atoms along the entire backbone -- 154 in the case of 1ABS. Figure \ref{fig8}d 
shows the thermal dependence of the helicity $H$.
Again, the result shows that there are three different phases in myoglobin as a function of $T$, matching 
those shown in the other panels in Figure \ref{fig8}. 

We also introduce the susceptibility $S_Q$ of a quantity $Q$, defined as follows, 
\begin{equation}
S_Q = \langle Q^2\rangle  - \langle Q \rangle^2 \,.
\label{SQ}
\end{equation}
In the thermodynamic limit of a second-order phase transition, the susceptibility diverges at the critical 
temperature. However, in a finite system, this divergence is replaced by a peak located near the putative 
transition point. If the maximum value of the peak does not increase with the volume, it signals the 
presence of a smooth crossover transition, while a peak that grows in proportion to the volume of the system 
is a characteristic of a discontinuous, first-order phase transition. A second-order transition then lies in 
between these two extremes.

Figures~\ref{fig8}c and \ref{fig8}d show the temperature evolution of both the radius of gyration 
susceptibility $S_{R_g}$ and the helicity susceptibility $S_H$. Both identify the three distinct phases of 
myoglobin, but their behaviors at the transition regimes are different: In the case of $S_{R_g}$ shown in 
Figure \ref{fig8}c, there is no peak.  Thus, either the transitions are not of second order, or the radius of gyration 
$R_g$ does not couple to critical fluctuations. In the case of helicity susceptibility $S_H$, we observe (slight) 
peaks at both transition regions, implying that the respective thermodynamic transitions do take place. Thus, 
in terms of susceptibility, the fluctuations in helicity better reflect the difference in the nature of the phases 
than the radius of gyration.

\subsection{Local topology and perestroikas in myoglobin}

The radius of gyration and the helicity are both geometrically determined order parameters.  Neither 
can reveal the role of local topology and perestroikas in the transitions between the different 
phases that we have displayed in Figures \ref{fig8}. To disclose how thermal dynamics can affect the local 
backbone topology, we start with two examples, both of them focusing on the local topology around 
the F-helix that is located between sites 86 and 95. We have chosen this segment since the F-helix is 
presumed to have an important role in ligand entry and exit. Moreover, the F-helix is known to be the
first helix that is 
affected when the temperature starts increasing, which is also observed in the Glauber dynamics 
simulations of (\ref{Esol2}) in \cite{Peng-2016, Begun-2019}. 

\begin{figure}[h]
        \centering
                \includegraphics[width=0.48\textwidth]{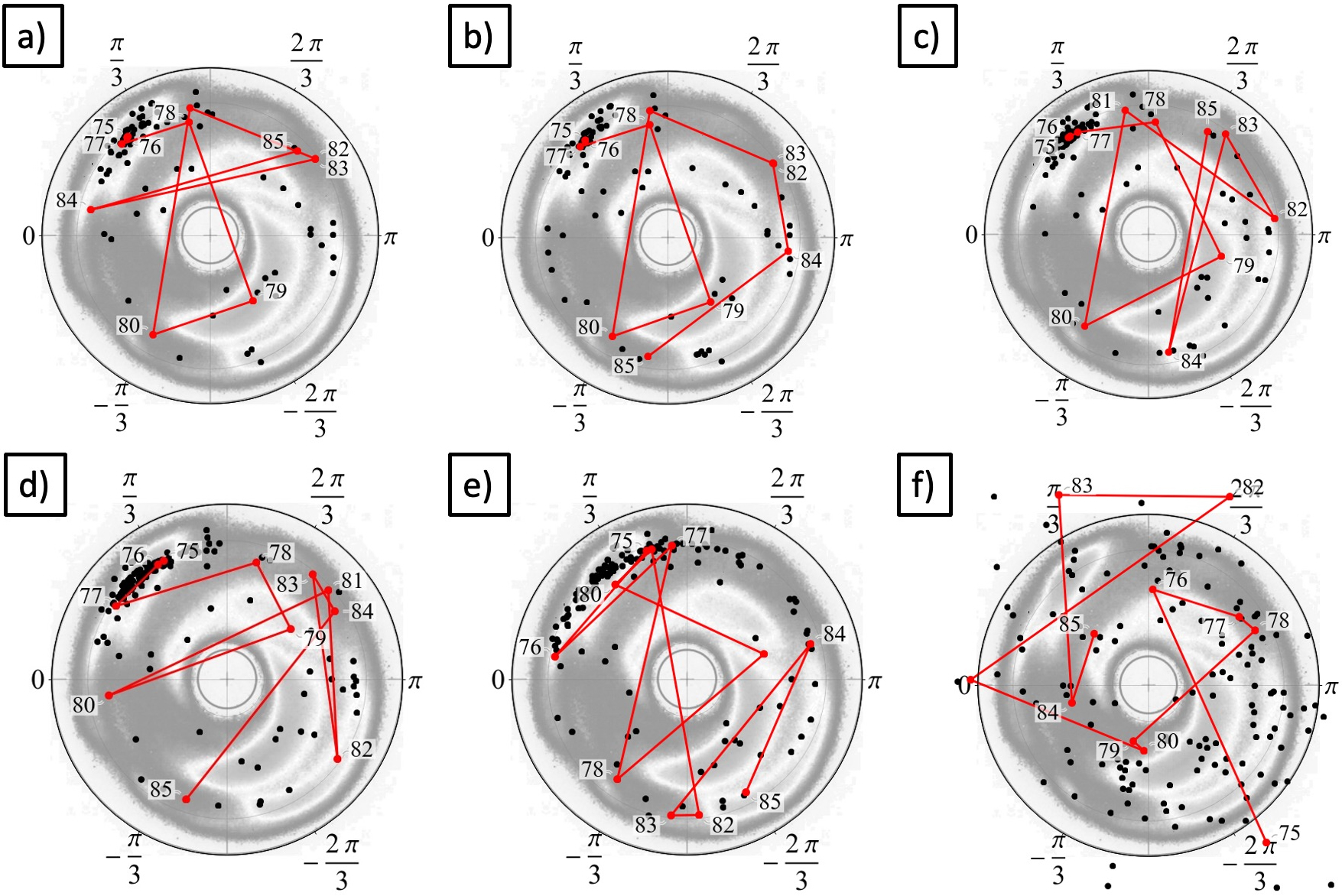}
        \caption{{
     Instantaneous snapshots of the trajectory 75-85 at different Glauber
     temperature factor values. In panel a) $T \approx 10^{-13}$, in panel b) $T \approx 10^{-12}$, in
      panel c) $T \approx 10^{-10}$, 
     in panel d) $T \approx 10^{-9}$,  in panel e) $T \approx 10^{-8}$, and  in panel f) $T \approx 10^{-5}$.
            }}
       \label{fig9}
\end{figure}

In Figure \ref{fig7}a, we have already presented the loop segment 75-85 on the annulus $\mathbb A$ 
in the case of the crystallographic 1ABS. In Figure \ref{fig9}a-f, we show, in terms of instantaneous 
snapshots along a Glauber dynamics trajectory, how this loop segment evolves on $\mathbb A$ as a function of $T$.  
%
%
%
%
%
%
%
%
%
%
%
%
%
%
%
%
%
%
%
%
%
%
%

It should be kept in mind that Figures~\ref{fig9} are instantaneous snapshots along a thermal 
trajectory. As such, the results we present are subject to thermal fluctuations with amplitudes that 
increase with increasing $T$.  While the changes in local topology are representative, the 
geometric details are not. 

Starting from the crystallographic structure, the $T$-value where we first start observing clearly 
visible temperature effects is $T \approx 10^{-13}$, shown in Figure \ref{fig9}a. This $T$-value 
corresponds to the end of the molten globule phase in terms of the geometric order parameters of 
Figure \ref{fig8}. In the snapshot Figure \ref{fig9}a, the vertex 84 has moved from the vicinity of the 
$\alpha$-helical region to the vicinity of the flattening line, and the vertex 85 has similarly moved 
from the vicinity of $\alpha$-helical region to $\tau \approx 2\pi/3$. There is no perestroika, and the 
local topology has remained intact. But since these thermal changes in the C$\alpha$-geometry 
occur in the vertices that are proximal to the F-helix,  this helix is starting to melt. This is in line with 
Figure \ref{fig8}d showing that helicity starts diminishing for $T$-values above $T \approx 10^{-14}$. 

Otherwise, there are only minor, local geometry changing movements in the remaining vertices along the 
backbone, as can be seen by comparing the placement of black dots in Figures \ref{fig7}a and \ref{fig9}a, 
denoting the remaining C$\alpha$ atoms of myoglobin. Thus, the original shape of the crystallographic 
structure remains largely intact.

In Figure~\ref{fig9}b, the temperature factor has increased to $T \approx 10^{-12}$, which is in the 
middle of the molten globule phase, according to Figures~\ref{fig8}. Except for the vertices 84 and 85,
 located right before the F-helix, the loop region between E and F helices remains largely intact. But
  there is now a branch point along the link connecting vertices 83 and 84, and the flattening 
  line needs to be crossed between vertex 85 and the $\alpha$-helical region. This increases the 
  folding index by +2, and we conclude that inflection point perestroikas,  in combination with either a 
  bi-flattening or a bi-branching perestroikas, can take place. The value of the folding index starts fluctuating.

In Figure \ref{fig9}c, we have a snap-shot at $T\approx 10^{-10}$, which corresponds to the 
beginning of the transition regime between molten globule and self-avoiding random walk phases. 
Additional perestroikas have taken place, affecting both the folding index between helices E and F 
and the positions of the vertices 84 and 85. We also observe an increase in backbone vertices (black dots in the figure) 
in the quadrant between the branch cut and $\tau = - \pi/2$, which proposes that perestroikas are indeed 
more common. There is also a slight widening in the geometry of vertices around the $\alpha$-helical region, 
showing that at these $T$-values, there is an onset of rapid decrease in helicity, consistent with Figure \ref{fig8}d.  
But by and large, the distribution of vertices on the annulus $\mathbb A$ in Figures \ref{fig9}a and \ref{fig9}c are 
still quite similar, suggesting that the changes in the overall shape of 
the backbone remain minor.

Figure \ref{fig9}d is a snap-shot at $T\approx10^{-9}$, in the transition region between molten globule and 
self-avoiding random walk phases. We observe the presence of several perestroikas that affect the structure 
of the loop, and the spreading of vertices around the $\alpha$-helical region continues, including those next 
to the E-helix. This is consistent with the increasing melting of helical regions, also observed in Figure \ref{fig8}d. 

Figure \ref{fig9}e is a snap-shot at $T\approx10^{-8}$, which is the low-temperature limit of the self-avoiding 
random walk phase, according to Figures \ref{fig8}. The melting of the loop close to the E-helix has continued, 
and closer toward the F-helix, there are several additional perestroikas, including inflection point perestroikas 
affecting the folding index. The vertices (black dots) are spreading more widely around the annulus $\mathbb A$. 

Finally, in Figure \ref{fig9}f, we have $T\approx10^{-4}$ corresponding to the self-avoiding random walk regime. 
The vertices are distributed quite randomly over the annulus $\mathbb A$, causing cascading  of perestroikas.

In summary, the examples in Figures \ref{fig9} suggest the following relation between the different phases 
and regimes in Figures \ref{fig8}, local backbone topology, and its perestroikas.

\vskip 0.2cm
$\bullet$ In the transition regime between collapsed and molten globule phases, the thermal fluctuations of the 
individual vertices start increasing, and the backbone chains swell; the changes are geometric with perestroikas 
rarely occurring, so that the shape of the backbone remains largely intact. 

\vskip 0.2cm
$\bullet$ In the molten globule phase, helices start melting, and local topology occasionally changes by 
occasional perestroikas.

\vskip 0.2cm
$\bullet$ During the transition from molten globule to self-avoiding random walk, the perestroikas become more
 common and start cascading. This causes the melting of helices and a
decay of super-secondary structures. 

\vskip 0.2cm
$\bullet$ The self-avoiding random walk phase is dominated
by frequent, cascading perestroikas and an increasingly random distribution of the bond and torsion angles on 
the annulus $\mathbb A$.

\subsection{Order parameters for local topology and perestroikas}

The previous example describes how temperature-driven transitions between the different phases 
correlate with the frequency of perestroikas, changing the number of flattening and branch points 
and the folding index along 
the backbone trajectory on $\mathbb A$. We now  introduce a novel order parameter, appropriate for 
estimating the $T$-evolution of perestroikas in terms of local topology changes. We start with the 
observation that whenever either bi-flattening or bi-branching perestroika takes place, there is an 
accompanied change in the sign of the torsion angle at a corresponding vertex. Thus, we assign 
the following quantity to each link, connecting a pair of neighboring vertices along the backbone,
\begin{equation}
b_i \ = \  
\begin{cases}
   \  \ 1 \hspace{0.7cm} {\rm if} \ \ \  \tau_i \tau_{i+1} \leq 0 \ \\ \ \  0 
   \hspace{0.7cm} {\rm if} \ \  \tau_i \tau_{i+1} >  0 
\end{cases}
\label{bival}
\end{equation}
and we define the following correlation function with 
$N=154$ in the case of 1ABS 
\begin{equation}
C(k) \ = \ \frac{1}{N-k} \ \left<\sum\limits_{i=1}^{N-k} b_i b_{i+k}\right> 
\label{Cs}
\end{equation}
where the average is taken over the thermal ensemble. 
Note that only links that cross either the flattening line or the branch cut can contribute to (\ref{Cs}). 
In particular, when $b_i$ and $b_{i+k}$ are both located along 
the same regular secondary structure, such as an $\alpha$-helix, 
there is no contribution from $b_i b_{i+k}$ to (\ref{Cs}).  

\begin{figure}[h]
        \centering
                \includegraphics[width=0.48\textwidth]{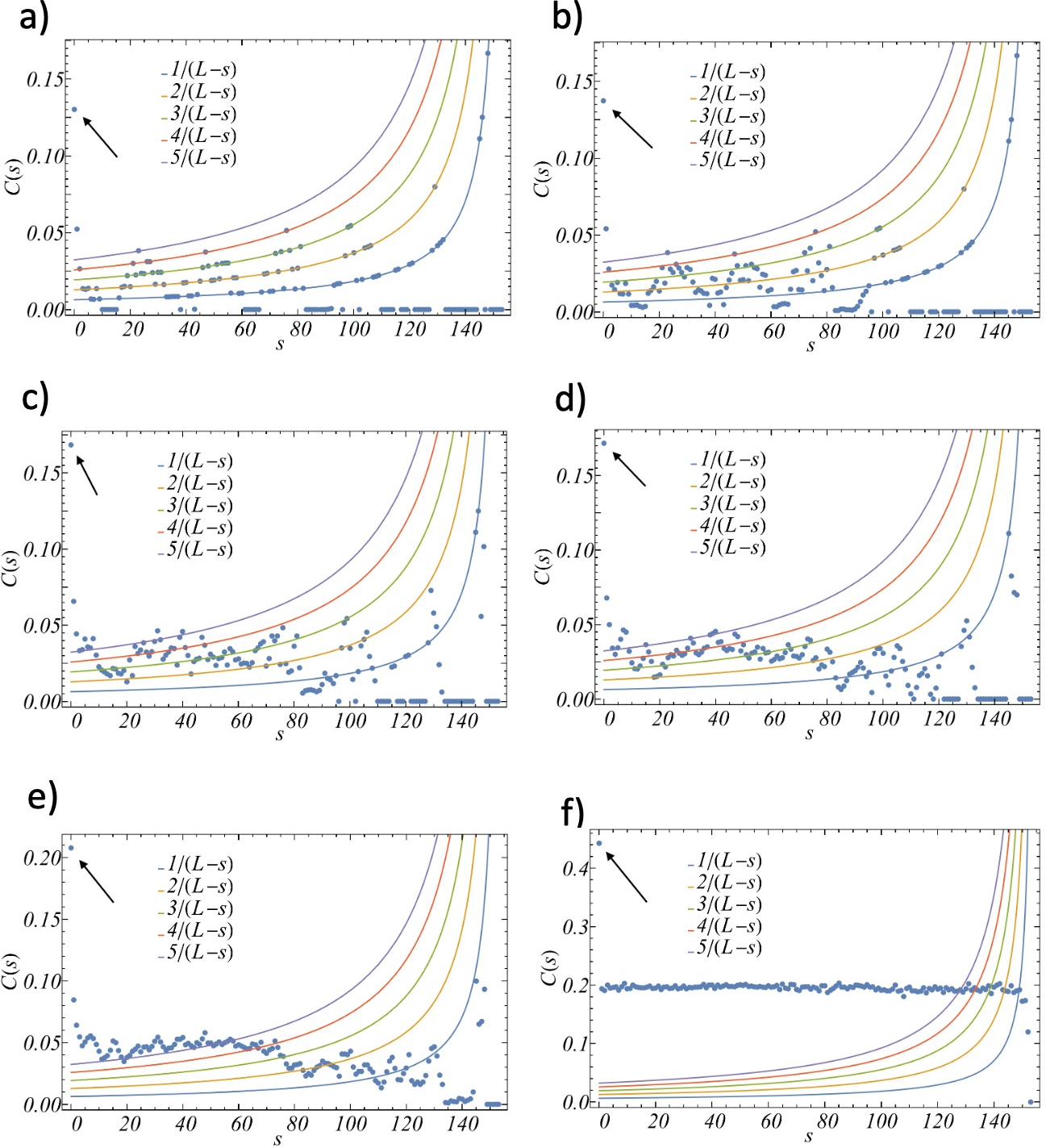}
%
%
%
%
%
%
%
%
%
%
%
        \caption{{
     Temperature factor dependence of the correlation function (\ref{Cs}). In panel a) $T\approx 10^{-14}$,
     in panel b) $T\approx 10^{-13}$, in panel c) $T\approx 10^{-11}$, in panel d) $T\approx10^{-9}$, in 
     panel e) $T\approx 10^{-8}$, in
      panel f) $T\approx 10^{-5}$. The arrows show the value of $C(0)$. All values are the averages over all 
      values in the ensemble, at the given temperature.
            }}
       \label{fig10}
\end{figure}

The Figure \ref{fig10} shows the $T$-dependence of $C(k)$  for several different $T$-values, extending over the
same range as Figure \ref{fig9}.
%
%
%
%
%
%
%
%
%
%
%
%
%
%
%
%
%
%
%
%
%
%
%
%
%
%
%
In the collapsed phase up until $T \approx 10^{-14}$ that we show in Figure \ref{fig10}a, we observe no 
change in 
$C(k)$; note that this is also the value of $T$ where, according to Figure \ref{fig7}a, the backbone enters the 
transition regime between the collapsed phase and molten globule. Remarkably, for these low-$T$ values, the 
values of $C(k)$  oscillate between the following level curves
\begin{equation} 
F_n(s) = \frac{n}{N-s}
\label{curvC}
\end{equation}
The levels $n=0,\dots,5$ are observed, with only one entry for $n=5$ in the case of 1ABS.  Furthermore, 
the contributions to the level $n=0$ come primarily from helices and, more generally, from regular secondary 
structures and segments that do not contain any flattening or branching points. 

The only exception to the level sets (\ref{curvC}) is the value of $C(0)$ that we identify with an arrow in all the 
Figures \ref{fig10}. The value of $C(0)$ counts the number of those links along the backbone that cross either 
the flattening line or the branch cut. Thus, any change in the value of $C(0)$ is a direct measure of perestroikas 
taking place. Since $C(0)$ has the constant value $C(0)\approx 0.13$ for all values below $T \approx 10^{-14}$, 
bi-flattening and bi-branching perestroikas are practically absent for these low-$T$ values. 

When $T$ increases so that the backbone chain enters the molten globule phase, the $C(k)$ starts to have 
values between the level sets (\ref{curvC}). This feature is shown in Figure \ref{fig10}b with $T \approx 10^{-13}$.

Here, and in the sequel, the entries show the corresponding thermal average values. 

We observe that the $n=0$ level curve has become less occupied, proposing that the melting of helices is in
 progress. It is also notable that the value of $C(0)$ has slightly increased, from
$ C(0) \approx  0.130$ at $T \approx 10^{-14}$ to $C(0)\approx  0.138$ at $T\approx 10^{13}$ implying that perestroikas are 
starting to take place. 

When the temperature factor increases to $T \approx 10^{-11}$ shown in Figure \ref{fig10}c, which is the 
upper limit of the molten globule phase, $C(k)$ is no longer organized along the level curves, and both 
$n=0$ and $n=1$ level curves are empty except for large $k$ values.  The occupation of level $n=2$ has 
also become quite sparse for smaller values of $k$.  The values of $C(0)$ have also grown to $C(0)\approx 0.168$ 
so that perestroikas are occurring more frequently.  

In Figure \ref{fig10}d, we show the result for $T\approx 10^{-9}$, which is in the transition regime between the 
molten globule phase and the self-avoiding random walk phase. The values of $C(k)$ are starting to re-organize 
linearly around the value $\sim 0.025$, except for larger $k$-values where we still observe the remnants of the 
$n=0$ level. The value of $C(0)$ has also increased, but only slightly, to $C(0) \approx 0.172$ so that there is no 
significant increase in the frequency of perestroikas. 

Figure \ref{fig10}e shows the values of $C(k)$ for $T\approx 10^{-8}$, which places us at the beginning of the 
self-avoiding random walk phase. The values of $C(k)$ are increasingly approaching a constant value, close
 to $\sim 0.05$, except for the large-$k$ where we still observe remnants of the level structure. The value of $C(0)$ 
 grows somewhat more rapidly; the value is now around $\sim 0.21$. 

Finally, in Figure \ref{fig10}d, we are in the self-avoiding random walk phase. The $C(k)$ is now essentially constant 
valued, $C(k) \approx 0.2$ (for $s\not=0$), and there has been a rapid growth in frequency of perestroikas, and  
$C(0)$ has grown more rapidly to $C(0) \approx 0.44$. We propose that both of these values are close to the 
universal values for a self-avoiding random walk chain. Notably, in the case of the ideal random walk, we would 
expect the value $C(0) = 0.5$. This is because the probability for any vertex to have a positive value equals the 
probability for it to have a negative value so that, on average, every other bond will cross either the flattening line 
or the branch cut. By similar reasoning, we expect to get $C(k) = 1/4$ for $k>0$ in the random walk phase.  

\begin{figure}[h]
        \centering
                \includegraphics[width=0.48\textwidth]{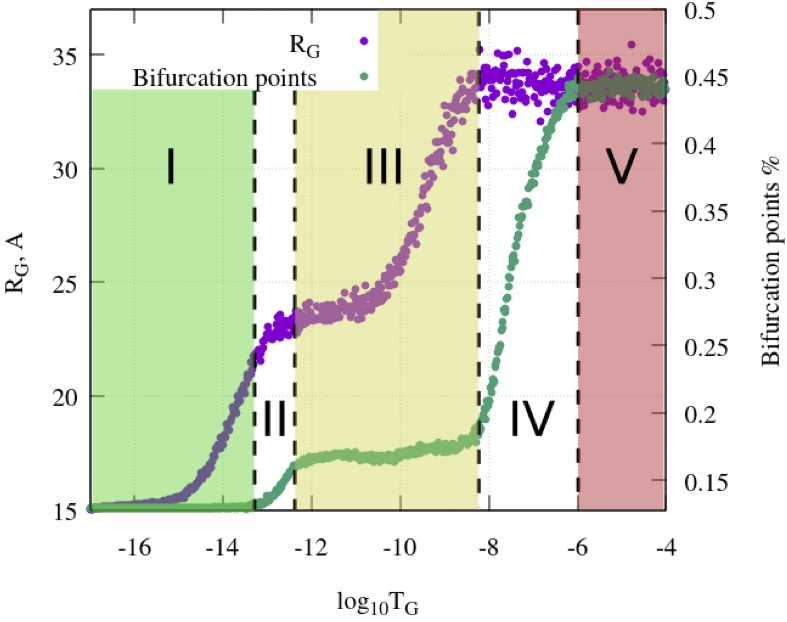}
%
%
%
%
%
%
%
%
%
%
%
%
%
%
%
%
%
%
        \caption{{a) Comparison of the radius of gyration $R_g$ (purple) and $C(0)$ (green), which counts the 
        relative number of flattening and branch cut points as a function of temperature factor $T$. 
        The labels I-V are the three phases (collapsed I, molten globule III, and SARW V) and the two 
        transition regimes (II and IV) as they are identified by $C(0)$. b) Binder cumulant of $C(0)$ (purple) and 
        susceptibility of $C(0)$ (green) as a function of $T$. The Binder cumulant clearly identifies the two 
        transition points, between the folded phase and the molten globule, and between the molten 
        globule and SARW, respectively.
            }}
       \label{fig11}
\end{figure}

Figure \ref{fig11}a summarizes the $T$-dependence of $C(0)$ in the different phases and transition regimes, 
comparing its value to the $T$-dependence of the radius of gyration.
%
%
%
%
%
%
%
%
%
%
%
%
%
%
%
%
%
%
%
%
%
%
%
%
%
The figure demonstrates how both the phases and the transitions between them are detected in a much more 
clear manner by the topological order parameter $C(0)$ for perestroikas, 
than by geometric order parameters such as $R_g$:  The three phases become much more
clearly recognizable, with transitions between them much more abrupt than in terms of $R_g$. The  transitions 
are also occurring at somewhat higher values of $T$. 
Indeed, since $C(0)$ has a topological origin  unlike $R_g$ it is not sensitive to changes such as swelling that 
affect only geometrical details, while leaving the topology,  {\it i.e.} the overall shape of the protein, largely intact. 

In Figure \ref{fig11}b we show the $T$-dependence of both the Binder cumulant (\ref{BQ}) and the susceptibility (\ref{SQ})
of $C(0)$. The Binder cumulant, in particular, marks the phase transition between the folded state and the molten 
globule, and the transition between the molten globule and the self-avoiding random walk phase in a visible manner, 
much more clearly than in the case of $R_g$ shown in Figure \ref{fig8}b. It displays two clear peaks. The first peak 
at around $T\approx 10^{-13}$ coincides with the decay of the level structures (\ref{curvC}) as shown in 
Figure \ref{fig10}b and the second peak at around $T\approx 10^{-8}$ coincides with the process of
 re-arrangement of the correlation function along a constant value as shown in Figure \ref{fig11}b. Notice that the 
 positions of the second peaks in susceptibility and the Binder cumulant do not coincide. This is a characteristic 
 behavior of a so-called pseudo-critical transition, in the case of a statistical finite-volume system.

By combining the information from $C(0)$ and $R_g$ and their Binder cumulants and susceptibilities, {\it i.e.} 
topological and geometrical information, we obtain a more complete, refined picture of the thermal evolution 
with the three-phase and their transition regimes clearly identified.  For example,  the transition between the 
collapsed phase to the molten globule phase starts with an initial swelling of the backbone that is detected by 
$R_g$. When $R_g$ reaches a value that corresponds to the molten globule, there is a rapid change 
in the frequency of perestroikas that change the local topology, as described by $C(0)$ and its Binder cumulant. 
Remarkably, within the molten globule phase, the frequency of perestroikas barely changes.

Similarly, the transition from molten globule to SARW is initiated by an increase in the swelling, with local 
topology largely intact. When the radius of gyration has reached the value that characterizes the SARW phase, 
there is a topological transition with a large increase in the value of $C(0)$ signaling a phase transition due to
 cascading perestroikas; in this case, the Binder cumulant detects the transition as the inception of break-up 
 in the molten globule phase.

\section{Summary}

We have introduced the concept of local topology into protein research and we have developed the 
ensuing bifurcation theory as a methodology to understand and  describe the phase structure and thermally driven dynamics of 
protein C$\alpha$ backbones. To achieve this, we have adapted and expanded Arnol'd's perestroikas to 
create a framework suitable for discrete piecewise linear chains, such as the C$\alpha$ backbone. As a 
specific example, we have investigated myoglobin, modelling it as a topological multi-soliton solution to a 
discretized nonlinear Schr\"odinger equation (DNLS). Unlike all-atom molecular dynamics, which aims to 
provide detailed geometrical descriptions of an entire protein with individual atom-level precision, our 
adapted approach focuses on the backbone's local topology, and how it changes.

We have combined the DNLS free energy with the Glauber algorithm to simulate the phase structure and thermal 
dynamics of myoglobin, at the level of local topology changes.  Our findings indicate that within a specific phase,
 the local topology of a C$\alpha$ backbone shows minimal dependence on its detailed geometry, such as the 
 precise positioning of individual atoms. Therefore, when concentrating on local topology and its alterations, 
 our results obtained from studying myoglobin are not limited to this protein alone but possess much broader 
 applicability to globular proteins.  

We have identified three perestroikas that are significant for C$\alpha$ backbones and their topological phase 
transitions: Bi-flattening perestroika, bi-branching perestroika, and inflection point perestroika. We have 
demonstrated how each contributes to the phase structures and the transitions between the different phases 
in the case of globular proteins. Specifically, each phase of a protein backbone possesses a distinctive local 
topology with a corresponding perestroika pattern.

We have introduced a novel correlation function to analyze the local topology of a C$\alpha$ backbone, 
enabling the detection of perestroikas and their cascading. By combining this correlation function with 
traditional geometric order parameters, such as the radius of gyration, we have developed more comprehensive 
analytical tools.  By utilizing both geometric and topological perspectives, our approach unveils new insights into 
the protein phase structure and thermal dynamics.

We are confident that the concepts of local topology and perestroikas, previously unexplored 
in protein research, will prove to be invaluable tools, even more widely, in the physical analysis of string-like objects.

\section{Acknowledgements}

AB and AJN are supported by the Carl Trygger Foundation. AJN is also supported by the Swedish 
Research Council under Contract No. 2018-04411 and 2022-04037, by COST Action CA21109 (CaLISTA), 
and by COST Action CA211169 (DYNALIFE). AJN thanks Erwin Schr\"odinger International Institute for 
Mathematics and Physics for hospitality during the completion of this article. 
AM is supported by Grant No.FZNS-2024-0002 of the Ministry of Science and Higher Education of Russia.

\end{document}